\documentclass[10pt,pra,aps,amsmath,amssymb,twocolumn,superscriptaddress,hidelinks]{revtex4-2}

\usepackage{usepackages}

\makeatletter
\newcommand{\repeatablemaketitle}{%
  \@author@finish
  \title@column\titleblock@produce
  \suppressfloats[t]%
  \titlepage@sw{\vfil\clearpage}{}%
}
\makeatother

\makeatletter
\newcommand{\HideFromSupplementTOC}{%
  \let\SUPP@saved@section\l@section
  \let\SUPP@saved@subsection\l@subsection
  \let\SUPP@saved@subsubsection\l@subsubsection
  \let\l@section\@gobbletwo
  \let\l@subsection\@gobbletwo
  \let\l@subsubsection\@gobbletwo
}

\newcommand{\ShowInSupplementTOC}{%
  \let\l@section\SUPP@saved@section
  \let\l@subsection\SUPP@saved@subsection
  \let\l@subsubsection\@gobbletwo
}
\makeatother

\begin{document}

\title{A Superconducting Peierls Instability}

\author{P. Senarath Yapa\,\orcidlink{0000-0002-5031-4695}}
\email{pramodh.sy@gmail.com}
\affiliation{Department of Physics and Astronomy, Uppsala University, Box 524, 751 20 Uppsala, Sweden
}

\author{J. Maciejko\,\orcidlink{0000-0002-6946-1492}}
\affiliation{Department of Physics, University of Alberta, Edmonton, AB, Canada T6G~2E1}
\affiliation{Theoretical Physics Institute \& Quantum Horizons Alberta, University of Alberta, Edmonton, Alberta T6G 2E1, Canada}

\author{F. Marsiglio\,\orcidlink{0000-0003-0842-8645}}

\affiliation{Department of Physics, University of Alberta, Edmonton, AB, Canada T6G~2E1}
\affiliation{Theoretical Physics Institute \& Quantum Horizons Alberta, University of Alberta, Edmonton, Alberta T6G 2E1, Canada}

\author{A. M. Black-Schaffer\,\orcidlink{0000-0002-4726-5247}}
\affiliation{Department of Physics and Astronomy, Uppsala University, Box 524, 751 20 Uppsala, Sweden
}

\begin{abstract}
The Peierls instability is a foundational mechanism in condensed matter physics, showing how the electron--phonon interaction can transform a simple metal into an insulating state with a new lattice periodicity. In a one-dimensional (1D) metal the Peierls instability follows from the enhanced electronic response at wavevector $Q=2k_F$ (connecting the Fermi points) producing a Kohn anomaly: a softening of the phonon mode at the same wavevector. Condensation of this mode then generates the tell-tale Peierls periodic lattice distortion and charge-density wave (CDW) that gaps the electronic spectrum. 
Here we show an analogous instability at the edge of a two-dimensional (2D) superconductor, where a dispersing Andreev bound state (ABS) hosts Bogoliubov Fermi points at $\pm k_c$. The enhanced quasiparticle response at the connecting wavevector $Q=2k_c$ couples directly to pairing-fluctuations and produces a superconducting Kohn anomaly: a softening of a pairing mode at the same wavevector. Condensation of this mode then generates an edge pair-density wave (PDW) that gaps the ABS. We refer to this as a {\it superconducting Peierls instability} and identify the boundary quasiparticle structure that enables it. As a concrete realization, we consider a square-lattice extended Hubbard model whose mixed-symmetry $s+d+ip$ state hosts a dispersing ABS with zero-energy crossings at finite edge momenta. Using self-consistent Bogoliubov--de Gennes (BdG) calculations we show that the order parameter develops an edge PDW, with the wavevector set by the superconducting Kohn anomaly, that gaps the ABS crossings. Our results identify a novel mechanism for spontaneous translation-symmetry breaking in a superconductor. As this superconducting Peierls mechanism does not require an extensive zero-energy flat band or topologically protected edge states, it may apply broadly to unconventional superconductors.
\end{abstract}

\date{September 3, 2026}
\repeatablemaketitle

\addtocontents{toc}{\protect\HideFromSupplementTOC}

\section{Introduction}
\label{sec:introduction}
\begin{figure*}[!t]
  \centering
  \includegraphics[width=1\linewidth]{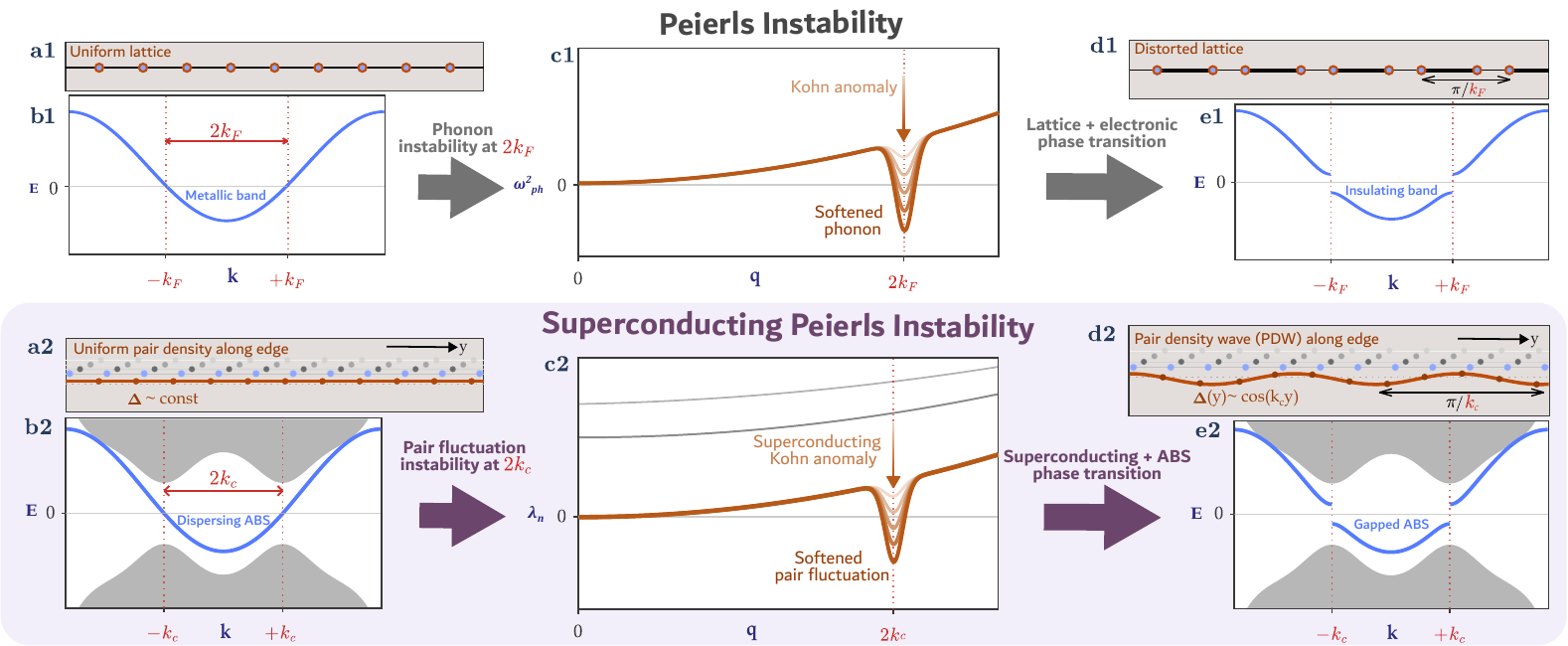}
  \caption{
Schematics of the conventional Peierls instability (top row) and its superconducting analog (bottom row). (a1, b1) Uniform 1D lattice and its electronic band, with Fermi points at $\pm k_F$. (c1) Kohn anomaly: progressive softening of the phonon mode at the nesting wavevector $q=2k_F$ driven by electron-phonon coupling. Colour gradient indicates increasing electron-phonon coupling. The instability occurs when $\omega_{\rm ph}^2$ becomes negative, i.e.~the phonon frequency becomes complex-valued. (d1, e1) Resulting distorted lattice with accompanying CDW and gapped electronic spectrum. (a2, b2) Translation-invariant edge of a semi-infinite 2D superconductor and its schematic BdG spectrum, including the bulk continuum and its Andreev bound-state (ABS) branch with counter-propagating zero-energy crossings at $\pm k_c$. (c2) Superconducting Kohn anomaly: suppression of an edge pairing-fluctuation stiffness at $q_y=2k_c$ driven by coupling to the ABS. The instability occurs when this stiffness becomes negative. (d2, e2) Resulting edge PDW and gapped ABS spectrum at $\pm k_c$. 
}
  \label{fig:roadmap}
\end{figure*}

The Peierls instability is the quintessential example of electronic structure driving a spontaneous reorganization of the underlying atomic lattice~\cite{Peierls1955}. In this mechanism, the Fermi surface selects a symmetry-breaking distortion of the underlying atomic lattice. In a one-dimensional (1D) metal, the Peierls instability is simplest and proceeds in three steps:
\begin{enumerate}[label=(\arabic*),leftmargin=*,itemsep=0pt,topsep=3pt]
\item The normal state has Fermi points at $\pm k_F$ with opposite group velocities.
\item The enhanced electronic susceptibility at the nesting wavevector $2k_F$ couples to the lattice through the electron--phonon interaction, producing a Kohn anomaly that softens the corresponding phonon mode~\cite{Kohn1959}.
\item The softened mode condenses, producing a periodic lattice distortion and charge-density wave (CDW) that gap the electronic spectrum at the Fermi points.
\end{enumerate}
This conventional Peierls instability is summarized in the top row of Fig.~\ref{fig:roadmap}, resulting in broken translation symmetry in the lattice and a metal-to-insulator transition. The general organizing principle is that a low-energy spectrum selects a finite ordering wavevector in another degree of freedom, and the resulting order removes the very states that generated the instability.

We show that this same sequence of events arises at the boundary of a two-dimensional (2D) superconductor. The role of the 1D metal is played by a dispersing Andreev bound state (ABS) band localized near the edge, while the role of the lattice is played by the superconducting order parameters. The superconducting Peierls instability thus proceeds in three steps:
\begin{enumerate}[label=(\arabic*),leftmargin=*,itemsep=0pt,topsep=3pt]
\item The uniform edge has Bogoliubov Fermi points at $\pm k_c$, corresponding to zero-energy ABS crossings with opposite group velocities.
\item The enhanced quasiparticle response at the connecting wavevector $2k_c$ couples to the superconducting order parameter(s), producing a superconducting Kohn anomaly that softens the corresponding pairing mode~\footnote{This is distinct from an earlier notion of a Kohn anomaly in a superconductor~\cite{Marsiglio1993PRBeosee, Egami1994PMSlehts,Aynajian2008Sci_egak,Johnston2011PRBcegk}, which refers to the renormalization of a bulk phonon's nesting-driven softening by the opening of the superconducting gap. We instead refer to the superconducting Kohn anomaly as a softening of the pairing-fluctuation stiffness itself, driven by BdG quasiparticle scattering between edge Andreev bound states.}.
\item The softened mode condenses, producing an edge pair-density wave (PDW)~\cite{Berg2009StripedSC,Agterberg2020PDWReview} that gaps the ABS spectrum at the zero-energy crossings.
\end{enumerate}
\nocite{Marsiglio1993PRBeosee,Egami1994PMSlehts,Aynajian2008Sci_egak,Johnston2011PRBcegk}
This superconducting Peierls instability is summarized in the bottom row of Fig.~\ref{fig:roadmap}, resulting in broken translation symmetry through the superconducting order parameter(s) and a gapped ABS spectrum.

The resulting edge PDW belongs to a broader class of boundary instabilities in which low-energy Andreev or Majorana modes are removed by spontaneously generated order. Previous examples include time-reversal breaking in $d$-wave superconductors~\cite{FogelstromPRL1997tic,CovingtonPRL1998oos,Rainer1998ABSReview,Huck1997TRSB,Higashitani1997PME,Miyawaki2015PRB_fse} and topological superconductors and superfluids~\cite{Grover2014BoundarySUSY,Park2015SurfaceMajorana,Li2017EdgeSUSY}, phase-crystal states at pair-breaking boundaries~\cite{Vorontsov2009BrokenSymmetryFilms,Holmvall2018BrokenTranslationEdges,Holmvall2020PhaseCrystals,WallWennerdal2020PhaseCrystal,Chakraborty2022DisorderRobustPhaseCrystal,Seja2025ImpurityPhaseCrystal}, spatially modulated states in confined superfluid $^3$He~\cite{VorontsovSauls2007He3Stripe,Levitin2019He3Modulated,Shook2020He3PDW,Yapa2022He3TriangularPDW}, edge ferromagnetism driven by ABS~\cite{Seja2024SurfaceSplitting,Matsubara2020PRB_eodip}, and interaction-driven Majorana-flat-band instabilities~\cite{Potter2014effm,Varada2026_nss,Hofmann2016PRB_eio,Li_2013}. In fact, Ryu and Hatsugai~\cite{Ryu2002PRL_too} have noted the possibility of a Peierls-like instability of a zero-energy flat band in a $d$-wave superconductor. However, all these examples are due to an edge flat band and thereby explicitly depend on the large DOS associated with an extensive set of zero-energy boundary modes. By contrast, the Peierls instability presented here is driven by isolated zero-energy crossings of a dispersive boundary band, whose momentum separation fixes the ordering wavevector in direct analogy with the conventional Peierls instability.

Replacing the flat-band requirement with a finite-wavevector nesting condition makes the superconducting Peierls mechanism applicable to a broader range of superconductors. The minimal requirement is only a finite nesting wavevector connecting counter-propagating zero-energy Bogoliubov states. At the 1D edge considered here, these are Bogoliubov Fermi points formed by zero-energy ABS crossings with opposite group velocities. Such nesting allows finite-$Q$ scattering to connect occupied and empty quasiparticle states near zero energy, thereby enhancing the pairing response. 
Pair-breaking boundaries of unconventional superconductors provide a natural setting for this structure. At such edges, reflection can connect trajectories that sample different signs or phases of the order parameter, allowing a dispersive ABS band to cross zero energy at finite edge momenta~\cite{Hu1994MGS,Lofwander2001ABS}. More generally, the same kinematic condition arises in the bulk when a finite wavevector connects zero-energy regions of a Bogoliubov Fermi surface~\cite{Agterberg2017BFS,Brydon2018BFS}. Here, we focus on the boundary realization, leaving possible bulk superconducting Peierls instabilities for future study.

As a concrete realization of the superconducting Peierls mechanism, we consider superconductivity in the square-lattice extended Hubbard model. Even with only onsite and nearest-neighbour interactions, mean-field studies suggest this model supports stable low-temperature mixed-symmetry phases~\cite{Nayak_2018,Hutchinson_2020,Hutchinson2019,SenarathYapa2025PC_mss,PramodhPhDThesis}. More generally, such coexistence of pairing components is a common feature of multichannel superconductivity~\cite{Annett1990,ODonovan1995,Musaelian1996, SigristUeda1991,Sigrist2005,NORMAND1994,betouras1997,Mitra1998, ANGILELLA2001,BASU2006,Medvedev2012,Timirgazin2019,Akbar2024,Sutradhar2024,Kheirkhah2020,Yerzhakov2026arxiv_iop}. We focus on the prevalent $s+d+ip$ phase, where spin-singlet and spin-triplet components coexist and both time-reversal and parity symmetries are spontaneously broken in the bulk. In a ribbon geometry, pair-breaking at the edge leads to a pair of counter-propagating ABS crossings over a broad range of interaction strengths. Imposing translation-invariance along the ribbon edge, ABS crossings occur at $\pm k_c$ and we find that an enhanced quasiparticle response leads to the static edge pairing-fluctuation kernel developing a finite-momentum suppression exactly at the connecting wavevector, $2k_c$. When the translational invariance along the edge is relaxed in self-consistent BdG calculations, an edge PDW develops at $Q_{\rm PDW}=2k_c$ and gaps the ABS. The coincidence of these three signatures identifies the PDW boundary modulation as a superconducting Peierls instability.
\section{Conventional Peierls Instability}
\label{sec:peierls_instability}

We first recall in more detail the conventional Peierls mechanism in a form that makes the superconducting analogy transparent. A 1D metal with Fermi wavevector $k_F$ is unstable at low temperature to a lattice distortion with wavevector $2k_F$, which opens a gap at the Fermi points [Fig.~\ref{fig:roadmap}(a1)--(e1)]. The instability is driven by the static electronic susceptibility, or Lindhard function~\cite{Mihaila2011arxiv_lfo},
\begin{equation}
\chi_0(q)
=
\frac{1}{N}\sum_k
\frac{f(\epsilon_k-\mu)-f(\epsilon_{k+q}-\mu)}
{\epsilon_{k+q}-\epsilon_k},
\label{eq:lindhard_new}
\end{equation}
where $N$ is the number of momenta, $\epsilon_k$ is the electron dispersion, $\mu$ is the chemical potential, and $f(E)$ is the Fermi--Dirac function. In 1D, the zero-temperature susceptibility $\chi_0(q)$ has a logarithmic singularity at $q=2k_F$. This divergence is the root of the Kohn anomaly~\cite{Kohn1959}: it reflects the fact that the wavevector $2k_F$ connects the two Fermi points, so an electron near $+k_F$ can scatter to an empty state near $-k_F$ (and vice-versa) at vanishing energy cost. This enhanced electronic response renormalizes the phonon frequency. For an electron-phonon coupling $g$ and bare phonon frequency $\omega_{{\rm ph},0}$, the renormalized phonon frequency may be written schematically as~\cite{Gruner2018}
\begin{equation}
  \omega_{\rm ph}^2(q)
  =
  \omega_{{\rm ph},0}^2
  -
  g^2\chi_0(q).
  \label{eq:peierls_instability_new}
\end{equation}
When $g^2\chi_0(2k_F)$ exceeds $\omega_{{\rm ph},0}^2$, the phonon mode at $2k_F$ becomes imaginary and the lattice distorts. The same wavevector that gives an enhancement in the electronic response therefore becomes the ordering wavevector for the lattice. When this $2k_F$ distortion condenses, it hybridizes the two electronic Fermi points and produces a CDW,
\begin{equation}
  \hham_{\rm CDW}^{\rm MF}
  =
  \Delta_{\rm CDW}\,
  \hat c^\dagger_{k_F}\hat c_{-k_F}
  +
  \Delta_{\rm CDW}^\ast\,
  \hat c^\dagger_{-k_F}\hat c_{k_F},
  \label{eq:conventional_peierls_gap}
\end{equation}
where $\hat c_k^{\dagger}$ ($\hat c_k$) creates (annihilates) an electron with momentum $k$, and $\Delta_{\rm CDW}$ denotes the CDW-induced hybridization gap between the two Fermi points. Microscopically, $\Delta_{\rm CDW}$ is proportional to the static $2k_F$ lattice displacement, and self-consistency relates this displacement to the particle-hole expectation value,
\begin{equation}
  \Delta_{\rm CDW}
  \propto
  \left\langle
    \hat c^\dagger_{-k_F}\hat c_{k_F}
  \right\rangle .
  \label{eq:cdw_gap_mean_field}
\end{equation}

\section{Superconducting Peierls Instability}
\label{sec:sc_peierls_instability}

The superconducting Peierls analog replaces the 1D electron band by a dispersing ABS localized at an edge [Fig.~\ref{fig:roadmap}(a2)--(e2)]. When this ABS has counter-propagating zero-energy crossings at $\pm k_c$, these two momenta play the role of boundary Fermi points, and the analog of the Peierls nesting vector is $Q_{\rm ABS}=2k_c$.

Further, in the superconducting case, the softened mode is not a phonon, but a fluctuation of the superconducting order parameter, localized to the edge as the ABS is edge-localized. 
Denoting the symmetry-allowed fluctuations about the translation-invariant order parameter by $\Phi_\alpha$, a fluctuation with edge momentum $q_y$ may be written schematically as
\begin{equation}
  \Phi_\alpha(x,q_y)=a_\alpha(q_y)w(x),
  \label{eq:edge_fluctuation_ansatz}
\end{equation}
where $a_\alpha(q_y)$ is the amplitude of a symmetry-allowed pairing-fluctuation labelled by $\alpha$, while $w(x)$ describes its transverse localization to the edge. To determine whether the translation-invariant edge is stable against such pairing-fluctuations, we examine the corresponding change in the (mean-field) free energy. Since the uniform state is a stationary point, the leading change is quadratic in the fluctuation amplitudes,
\begin{equation}
\delta F^{(2)}=\frac{1}{2}\sum_{q_y,\alpha,\beta}a_\alpha^\ast(q_y)[\mathcal K]_{\alpha\beta}(q_y)a_\beta(q_y),
\label{eq:quadratic_edge_free_energy_new}
\end{equation}
with the static edge pairing-fluctuation kernel
\begin{equation}
\mathcal K(q_y)=\mathcal K^{(0)}(q_y)-\Pi(q_y).
\label{eq:sc_peierls_new}
\end{equation}
This kernel follows from the standard Gaussian pairing-fluctuation expansion of the fermionic effective action~\cite{AltlandSimons2010,Protter2021PRBfia}. The matrix $\mathcal K(q_y)$ is equivalently the inverse static pairing-fluctuation propagator restricted to the edge fluctuation subspace. The bare kernel $\mathcal K^{(0)}(q_y)$ describes the microscopic pairing cost of fluctuations in the superconducting condensate, while $\Pi(q_y)$ is the static BdG quasiparticle response. Equation~\eqref{eq:sc_peierls_new} is the superconducting counterpart of Eq.~\eqref{eq:peierls_instability_new} and $\Pi(q_y)$ plays the role of $g^2\chi_0(q)$, with the normal density matrix element replaced by the matrix element of a pairing-fluctuation (see Methods Sec.~\ref{app:edge_susceptibility} for details). Diagonalizing $\mathcal K(q_y)$ gives us the pairing-fluctuation stiffness eigenvalues, $\lambda_n(q_y)$.
We define the superconducting Kohn anomaly as the cusp-like suppression of at least one stiffness eigenvalue at $q_y=Q_{\rm ABS}$. If any eigenvalue then becomes negative, the translation-invariant edge becomes unstable to a static finite-momentum pairing deformation, an edge-localized PDW. 

Projected onto the low-energy ABS subspace, the finite-momentum pairing deformation contains the gap-opening term
\begin{equation}
\hham_{\rm PDW}^{\rm MF}\supset\Delta_{\rm PDW}\hat{\gamma}^\dagger_{k_c}\hat{\gamma}_{-k_c}+\Delta_{\rm PDW}^\ast\hat{\gamma}^\dagger_{-k_c}\hat{\gamma}_{k_c},
\label{eq:sc_peierls_abs_gap}
\end{equation}
where $\hat{\gamma}_{k}^{\dagger}$ ($\hat{\gamma}_{k}$) creates (annihilates) an edge-localized Bogoliubov quasiparticle with momentum $k$. The pairing deformation in the Bogoliubov basis also contains $\hat\gamma\hat\gamma$ and $\hat\gamma^\dagger\hat\gamma^\dagger$ terms, as discussed in the Supplementary Material (SM). Microscopically,
\begin{equation}
\Delta_{\rm PDW}\propto\left\langle\hat{\gamma}^\dagger_{-k_c}\hat{\gamma}_{k_c}\right\rangle\supset\left\langle\hat c^\dagger_{-k_c}\hat c^\dagger_{-k_c}\right\rangle, 
\label{eq:abs_gap_mean_field_main}
\end{equation}
where the first expectation value is expressed in terms of Bogoliubov quasiparticles and is the direct analog of the particle--hole expectation value in Eq.~\eqref{eq:cdw_gap_mean_field}. The second expectation value arises because Bogoliubov quasiparticles mix electron and hole operators, thereby generating an anomalous Cooper-pair amplitude with center-of-mass momentum $-2k_c$, with the Hermitian-conjugate term in Eq.~\eqref{eq:sc_peierls_abs_gap} supplying the corresponding $+2k_c$ component. Here, the electron operators are written schematically; see SM for the full derivation. Thus, the same finite-$Q$ mean field appears as a gap-opening hybridization in the ABS basis and as a PDW with components at $Q=\pm2k_c$ in the electronic basis, which completes the superconducting analog of the Peierls transition.

\section{Dispersing Andreev Bound-State Crossings}
\label{sec:general_abs_condition}

The superconducting Peierls mechanism relies on a dispersing edge ABS with zero-energy crossings at finite and opposite momenta. We now examine how such crossings arise at a specular superconducting boundary. Because momentum parallel to the edge is conserved, each boundary state can be viewed in terms of quasiparticle trajectories that approach and reflect from the surface. This perspective relates the ABS spectrum to the order parameters sampled along those trajectories and provides the starting point for identifying the boundary Bogoliubov Fermi points.

Within the Andreev approximation~\cite{Andreev1964}, we consider quasiparticle trajectories near the normal-state Fermi surface. For a specular boundary with conserved momentum $k_\parallel$, the normal-state Fermi surface fixes a perpendicular momentum $k_{F\perp}(k_\parallel)$, and reflection relates the two
trajectories,
\begin{equation}
  \bm{k}_\pm
  =
  \bigl(\pm k_{F\perp}(k_\parallel),k_\parallel\bigr).
\end{equation}
The edge spectrum is controlled by the order parameters sampled along these trajectories~\cite{Vorontsov2018PTRSA_abs},
\begin{equation}
  \Delta_\pm(k_\parallel)
  =
  \Delta\!\left(
    \pm k_{F\perp}(k_\parallel),k_\parallel
  \right).
  \label{eq:general_delta_pm_new}
\end{equation}

The boundary is pair-breaking when reflection connects trajectories that sample inequivalent pair potentials. In particular, a phase difference between $\Delta_+$ and $\Delta_-$ produces subgap ABS~\cite{KashiwayaTanaka2000}. 

We focus first on the common case in which specular reflection leaves the magnitude of the order parameters unchanged,
$|\Delta_+(k_\parallel)|=|\Delta_-(k_\parallel)|$. This condition is exact when the gap magnitude is invariant under reflection and provides a particularly transparent analytic form of the ABS spectrum. After removing the common phase by a global gauge transformation, the two projected order parameters can then always be written as
\begin{equation}
  \Delta_\pm(k_\parallel)
  =
  A(k_\parallel)\pm iB(k_\parallel),
  \qquad A,B\in\mathbb{R}.
  \label{eq:general_A_B_new}
\end{equation}
Thus, in this gauge, $A$ is the reflection-even component, while $iB$ is the reflection-odd component; this real-even/imaginary-odd form follows from the equal-magnitude condition. For this case of equal incoming and reflected gap magnitudes, the standard
specular-surface ABS result gives~\cite{Sauls2018PTRSA_abs}
\begin{equation}
  E_{\rm ABS}(k_\parallel)
  =
  \pm A(k_\parallel),
  \label{eq:general_abs_energy_new}
\end{equation}
as shown explicitly in the SM. Thus $A$ controls the ABS dispersion, while $B$ produces the phase difference under reflection and controls the localization of the bound state. The superconducting Peierls mechanism requires zero-energy crossings at $k_c$ such that $A(k_c)=0$ and $B(k_c)\neq 0$. In words, a reflection-even projected gap component must cross zero along the projected Fermi surface, while a reflection-odd component remains finite to provide the needed phase twist for a localized ABS.

In addition, the two (same-edge) crossings connected by $Q_{\rm ABS}=2k_c$ must be counter-propagating, i.e.~with opposite group velocities $v_{\rm ABS}(k_\parallel)$, where $ v_{\rm ABS}(k_\parallel)  =   \frac{\partial E_{\rm ABS}(k_\parallel)}{\partial k_\parallel}$.
This condition ensures that a finite-$Q$ fluctuation connects an occupied low-energy ABS state to an empty one. If the two crossings instead belong to a purely chiral branch with co-propagating states, the same scattering vector connects filled states to filled states, or empty states to empty states, and the Peierls enhancement is absent, see SM. To summarize, because sign-changing order parameters are a defining feature of unconventional superconductors, it should be possible to find systems with suitable edge orientations that satisfy these conditions. Notably, neither a zero-energy flat band nor topology is required.

\section{Square-Lattice Extended Hubbard Model}
\label{sec:model_geometry}
To illustrate the superconducting Peierls mechanism, we study the square-lattice extended Hubbard model with interactions,
\begin{equation}
\begin{aligned}
\hham = & -t \sum_{\langle i,j\rangle, \sigma} \left( \hat{c}_{i \sigma}^{\dagger} \hat{c}_{j \sigma} + \hat{c}_{j \sigma}^{\dagger} \hat{c}_{i \sigma} \right) -\mu \sum_{i, \sigma} \hat{n}_{i \sigma}\\
&+U \sum_i \hat{n}_{i \uparrow} \hat{n}_{i \downarrow} +V \sum_{\langle ij\rangle, \sigma,\sigma'} \hat{n}_{i \sigma} \hat{n}_{j \sigma'}.
\end{aligned}
\label{eq:extended_hubbard_new}
\end{equation}
Here $i$ and $j$ label lattice sites, $\langle i,j\rangle$ denotes nearest-neighbour pairs, $\sigma=\uparrow,\downarrow$ is the spin index, $t$ is the nearest-neighbour hopping, $\mu$ is the chemical potential, and $U$ and $V$ are onsite and nearest-neighbour interactions. 

A mean-field decoupling in the superconducting pairing channels allows onsite $s$-wave spin-singlet pairing ($\Delta_0$ order parameter) and nearest-neighbour bond pairing, resolved into extended-$s$-wave spin-singlet ($\Delta_{s^\ast}$), $d_{x^2-y^2}$ spin-singlet ($\Delta_d$), and $p$-wave spin-triplet components ($\Delta_{p_x}$ and $\Delta_{p_y}$). Details of the mean-field decoupling, resulting self-consistent BdG equations, and phase conventions are given in Methods Secs.~\ref{app:bdg_formalism} and \ref{app:op_conventions}.

\begin{figure}[!t]
  \centering
  \includegraphics[width=1\linewidth]{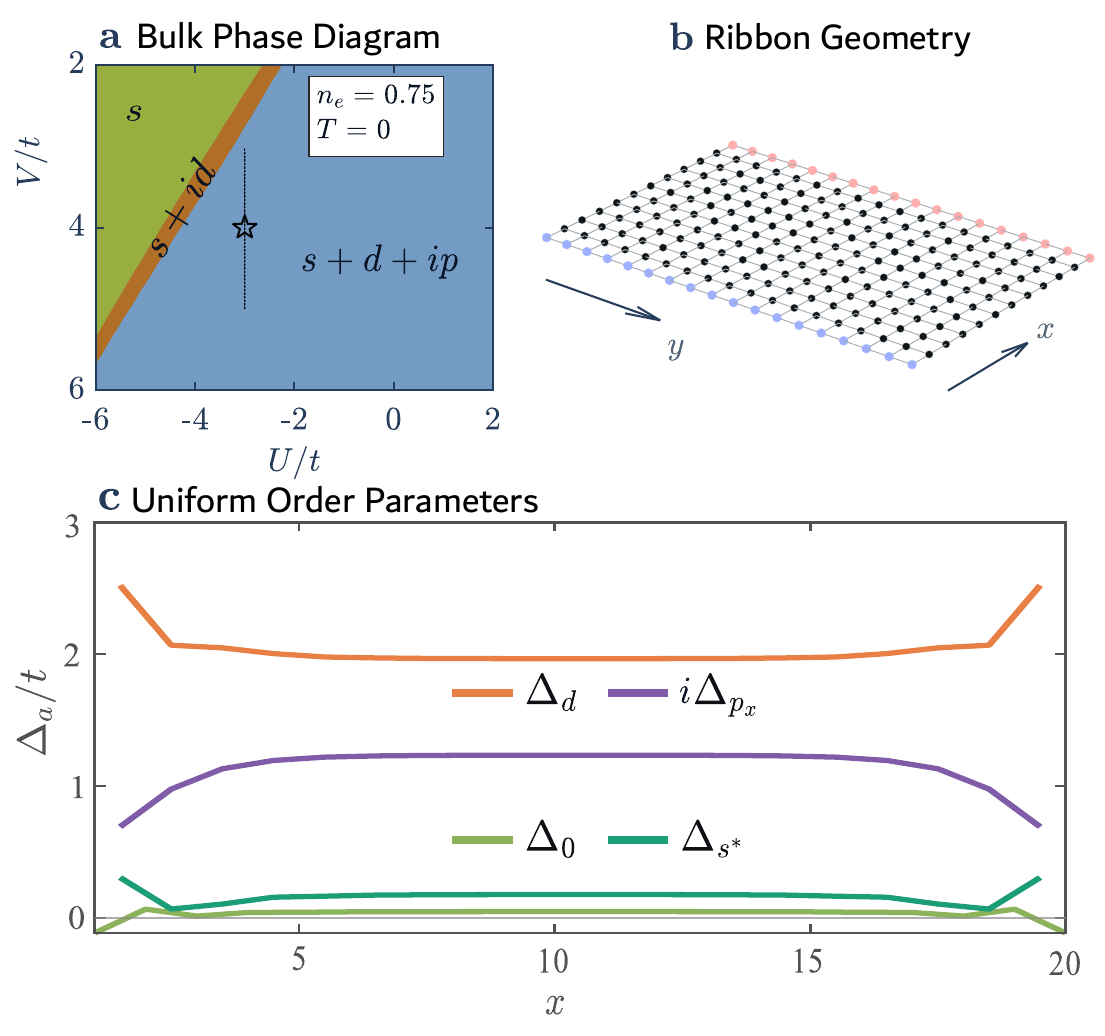}
  \caption{
Bulk phase diagram, ribbon geometry and the uniform ($y$-translation-invariant) order parameters on the ribbon. (a) Mean-field bulk phase diagram at $n_e=0.75$ and $T=0$ in the $(U/t,V/t)$ plane, showing $s$ (green), $s+id$ (orange), and $s+d+ip$ (blue) phases. Dashed vertical line corresponds to the cut examined in the SM, and {\small{\FiveStarOpen}} marks the $U/t=-3$, $V/t=-4$ point used in the main text. (b) Ribbon geometry with periodic boundary conditions along $y$, and open boundaries along $x$ (blue and pink open edges). (c) Uniform order parameter profiles solved using self-consistent BdG across an $N_x=20$ ribbon at {\small{\FiveStarOpen}}, enforcing translation invariance along $y$.
}
  \label{fig:fig2}
\end{figure}

Figure~\ref{fig:fig2}(a) shows the bulk zero-temperature self-consistent mean-field phase diagram at filling $n_e=0.75$. An $s+d+ip$ phase occupies a large region of the $(U/t,V/t)$ plane, consistent with previous studies~\cite{Nayak_2018,Hutchinson_2020, Hutchinson2019,SenarathYapa2025PC_mss,PramodhPhDThesis}. In the bulk, the $p_x$- and $p_y$-oriented mixed states are symmetry-related and degenerate. 
Open boundary conditions (OBC) distinguish the two degenerate bulk states as fully OBC self-consistent calculations show that both orientations develop an edge PDW, with the $s+d+ip_x$ ($s+d+ip_y$) state developing a PDW on edges normal to $x$ ($y$), see SM. Thus, we can here focus on the $s+d+ip_x$ phase in a ribbon geometry with an edge normal to $x$, see Fig.~\ref{fig:fig2}(b), where the edges are marked in colour.
We also focus on parameters $U/t=-3$, $V/t=-4$, marked by {\small{\FiveStarOpen}} in Fig.~\ref{fig:fig2}(a), to demonstrate the Peierls instability as a representative of a broader region.

\section{Andreev Bound States and Gap Opening}
\label{sec:parent_abs}
In order to provide the parent state from which the edge PDW instability develops, we first artificially
constrain the superconducting order parameters to remain translation invariant along the edge, i.e.~along the $y$-direction. We solve the resulting mean-field Bogoliubov--de Gennes (BdG) equations self-consistently, with $k_y$ as a good quantum number. The resulting real-space order-parameter profiles across the ribbon (along $x$) are shown in Fig.~\ref{fig:fig2}(c). The order parameters approach their uniform bulk values
toward the center of the ribbon, but some display pair-breaking at the edges.

\begin{figure}[!t]
  \centering
  \includegraphics[width=1\linewidth]{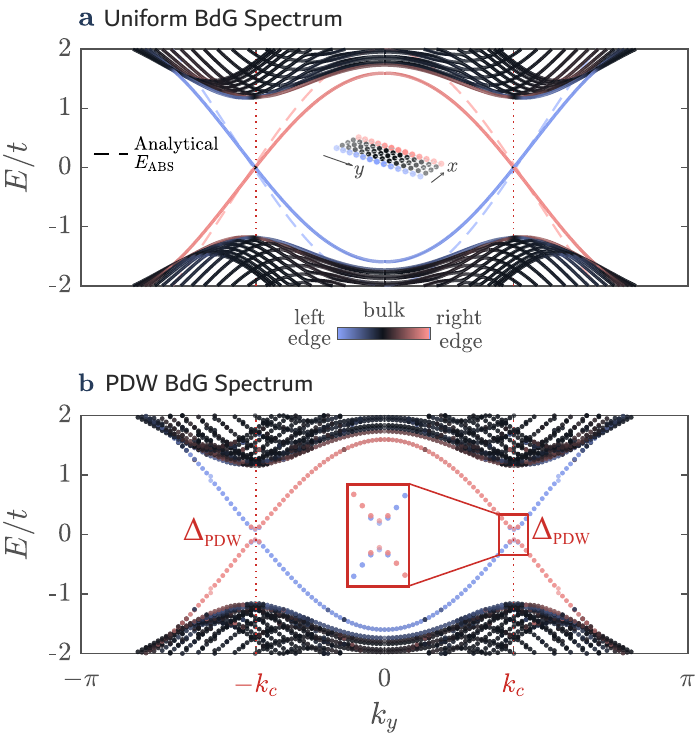}
  \caption{
ABS spectrum and gap opening in the edge PDW state. (a) BdG spectrum of the $y$-translation-invariant $N_x=20$, $N_k=600$ ribbon at $U/t=-3$, $V/t=-4$, and $n_e=0.75$. States are coloured according to their center-of-mass in the $x$ direction, with blue (red) indicating localization near the left (right) edge and dark colours indicating bulk-like states, see Methods Sec.~\ref{app:abs_tracking}. The ABS branches cross zero energy at $k_y=\pm k_c$ (red dotted vertical lines) with opposite velocities, where $k_c\simeq1.340$. Dashed curves show the analytic ABS dispersion $E_{\rm ABS}(k_y)=\pm A(k_y)$ (Eq.~\ref{eq:general_sdip_A}).
(b) Momentum-resolved BdG spectrum after translational invariance in $y$ is relaxed and the self-consistent edge PDW shown in Fig.~\ref{fig:fig4} forms. Since $k_y$ is no longer a good quantum number, the real-space BdG eigenstates are instead Fourier resolved in the momentum basis of the translation-invariant ribbon, see Methods Sec.~\ref{app:abs_tracking}. $\Delta_{\rm PDW}$ marks the gap opening at $k_c$, highlighted by the inset.
}
  \label{fig:fig3}
\end{figure}

The corresponding self-consistent BdG spectrum of the translation-invariant ribbon is
shown in Fig.~\ref{fig:fig3}(a). It contains edge-localized ABS branches
that cross zero energy at $k_y=\pm k_c$ with opposite velocities. 
For this $x$-normal edge, the momentum parallel to the boundary is
$k_\parallel=k_y$, while $k_\perp=k_x$.
Correspondingly, the Fermi-surface momentum introduced in
Sec.~\ref{sec:general_abs_condition} becomes
$k_{F\perp}(k_\parallel)=k_{Fx}(k_y)$. To also obtain an analytic estimate for the ABS dispersion of the translation-invariant parent state, we approximate the pair potential sampled by quasiparticles near the edges as

\begin{equation}
\begin{aligned}
  \Delta(\mathbf{k})
  =
  &\,
  \tilde{\Delta}_0
  +
  \frac{\tilde{\Delta}_{s^\ast}}{2}
  \left(
    \cos k_x+\cos k_y
  \right)
  \\
  &+
  \frac{\tilde{\Delta}_d}{2}
  \left(
    \cos k_x-\cos k_y
  \right)
  +
  i\tilde{\Delta}_{p_x}\sin k_x ,
\end{aligned}
\label{eq:general_sdip_gap}
\end{equation}
where the tildes denote the self-consistent order parameters evaluated near the boundary. This realizes the reflection-even/reflection-odd structure introduced in Sec.~\ref{sec:general_abs_condition}. We again consider quasiparticle trajectories near the normal-state Fermi surface. Reflection sends $k_x \rightarrow-k_x$, while leaving $k_y$
unchanged. The $p_x$ contribution thus changes sign and provides the
reflection-odd component,
\begin{equation}
  B(k_y)
  =
  \tilde{\Delta}_{p_x}
  \sin\!\bigl[k_{Fx}(k_y)\bigr],
\end{equation}
while the reflection-even component is
\begin{equation}
  A(k_y)
  =
  \tilde{\Delta}_0
  -
  \frac{\mu}{4t}
  \left(
    \tilde{\Delta}_{s^\ast}
    +
    \tilde{\Delta}_d
  \right)
  -
  \tilde{\Delta}_d\cos k_y .
  \label{eq:general_sdip_A}
\end{equation}
The $p_x$ component thus supplies phase difference under reflection, while the even-parity components determine the ABS dispersion, through the general result $E_{\rm ABS}(k_y)=\pm A(k_y)$. The full derivation is given in the SM.

\begin{figure*}[!t]
  \centering
  \includegraphics[width=1\linewidth]{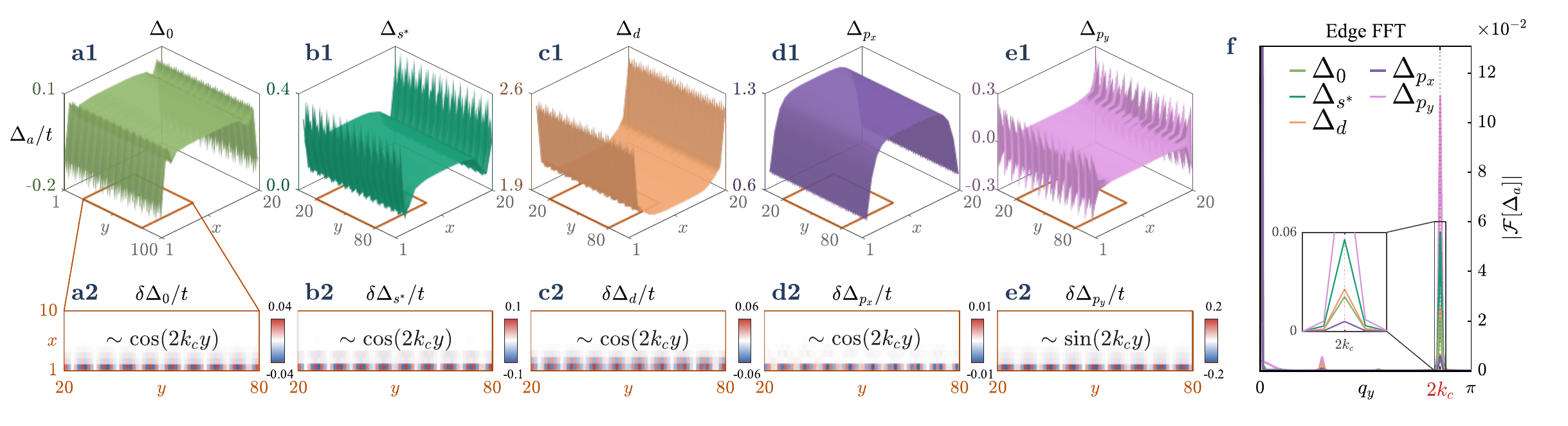}
  \caption{
Real-space structure of the edge PDW order parameters in ribbon geometry. (a1)--(e1) $\Delta_0$, $\Delta_{s^\ast}$, $\Delta_d$, $\Delta_{p_x}$, and $\Delta_{p_y}$ order parameters for an $N_x=20$, $N_y=150$ ribbon at $U/t=-3$, $V/t=-4$, and $n_e=0.75$. For clarity, the real-space panels display only 100 of the 150 lattice sites along $y$. Orange boxes indicate the edge regions isolated in (a2)--(e2) where the modulated component is isolated by subtracting the average of the order parameter along $y$ at each fixed $x$. The plotting and phase conventions are defined in Methods Sec.~\ref{app:op_conventions}. (f) Fast Fourier transform (FFT) of the order parameters at the edge, with the dominant peaks at $q_y=2k_c$ indicated by vertical red dotted line, with inset showing that every order parameter shares this Fourier peak.
}
  \label{fig:fig4}
\end{figure*}

The analytic dispersion $E_{\rm ABS}(k_y)=\pm A(k_y)$, plotted as dashed curves in Fig.~\ref{fig:fig3}(a),
is in excellent agreement with the numerical edge spectrum. Importantly,
these ABS branches do not connect the positive- and negative-energy bulk
continua across the superconducting gap. The absence of such bulk-connectivity is consistent with the states not being topologically protected, emphasizing that the Peierls mechanism does not require a topological edge state, but only dispersive ABS with zero-energy crossings. Such surface Andreev states are a standard consequence of sign- or phase-changing unconventional superconducting order parameters at reflecting boundaries.

Having identified counter-propagating ABS crossings, we next remove the constraint of translational invariance in $y$ and allow the order parameters to vary fully self-consistently along the edge. The resulting solution spontaneously breaks translation symmetry along the edge and notably opens a gap in the low-energy edge spectrum at the original zero-energy ABS crossings $k_c$. Figure~\ref{fig:fig3}(b) shows the corresponding momentum-resolved spectrum, which clearly displays gaps, indicated by $\Delta_{\rm PDW}$.

\section{Self-Consistent Edge PDW}
\label{sec:real_space_pdw}
We analyze the fully self-consistent order parameters in the ribbon geometry in Fig.~\ref{fig:fig4} responsible for the ABS gap opening. Deep in the bulk, the system remains in the uniform $s+d+ip_x$ phase, while near the open boundaries the order parameters modulate along $y$. The modulation is not confined to a single pairing channel: the onsite $s$, extended-$s$, $d$, and $p_x$ components all acquire edge oscillations, while a $p_y$ component, absent in the $y$-translation-invariant ribbon state, is also induced near the edge. 

The lower panels of Fig.~\ref{fig:fig4} isolate the modulated component by subtracting the $y$-averaged order-parameter profile at each fixed $x$. The resulting oscillations are localized near the boundary, decaying into the bulk and showing that the modulation is an edge PDW rather than a bulk PDW. The dominant components $\Delta_0$, $\Delta_{s^\ast}$, $\Delta_d$, and $\Delta_{p_x}$ all have $\cos(2k_c y)$-like oscillations, while the induced $\Delta_{p_y}$ component is shifted by $\pi/2$ and is approximately $\sin(2k_c y)$-like. This relative phase structure follows from the edge symmetry, see SM. 

We verify the period of the edge modulation by plotting the Fourier spectrum in Fig.~\ref{fig:fig4}(f). Its dominant nonzero peak occurs at $Q_{\rm PDW}=2k_c$, where $\pm k_c$ are the zero-energy ABS crossings. Thus the fully self-consistent solution indeed selects the same wavevector predicted by the superconducting Peierls mechanism. A direct comparison of the mean-field Helmholtz free energies confirms that the edge PDW is energetically favoured over the unmodulated $s+d+ip_x$ solution; see the SM.

\section{Superconducting Kohn Anomaly}
\label{sec:edge_susceptibility}

Having found that the edge PDW selects exactly the wavevector $Q_{\rm PDW}=2k_c$, set by the ABS zero-energy crossings, we now test the defining signature of the Peierls mechanism: whether the $y$-translation-invariant $s+d+ip_x$ reference edge exhibits a Kohn-anomaly-like softening of a pairing mode at the same wavevector. For this, we use the edge-fluctuation basis introduced in Eq.~\eqref{eq:edge_fluctuation_ansatz} and evaluate the stiffness matrix $[\mathcal K]_{\alpha\beta}(q_y)$ defined in Eqs.~\eqref{eq:quadratic_edge_free_energy_new} and \eqref{eq:sc_peierls_new}. For the $s+d+ip_x$ ribbon, the fluctuation indices $\alpha$ and $\beta$ run over the real and imaginary components of the five order parameters $\Delta_0$, $\Delta_{s^\ast}$, $\Delta_d$, $\Delta_{p_x}$, and $\Delta_{p_y}$. The $\lambda_n(q_y)$ eigenvalues of $\mathcal K(q_y)=\mathcal K^{(0)}(q_y)-\Pi(q_y)$ give the stiffnesses of the allowed pairing modes. A superconducting Kohn anomaly appears when the quasiparticle response $\Pi(q_y)$ strongly suppresses one of these eigenvalues at the wavevector connecting the low-energy ABS states.

The key property for the Kohn anomaly is thus the quasiparticle response $\Pi(q_y)$, which we compute from the BdG spectrum of the translation-invariant ribbon. This response renormalizes the pairing stiffness in direct analogy with the electronic renormalization of the phonon mode in the conventional Peierls problem. Explicitly,
\begin{equation}
\begin{aligned}
\Pi_{\alpha\beta}(q_y)
&=
\frac{1}{2N_y}
\sum_{k_y,m,n}
M_\beta^{mn}\left[M_\alpha^{mn}\right]^\ast
\\
&\quad\times
\frac{
f(E_n(k_-))-f(E_m(k_+))
}{
E_m(k_+)-E_n(k_-)
},
\end{aligned}
\label{eq:main_pi_spectral}
\end{equation}
where $k_\pm=k_y\pm q_y/2$ and
\begin{equation}
M_\alpha^{mn}
=
\langle m,k_+|
\Lambda_\alpha^{\rm edge}(k_y,q_y)
|n,k_-\rangle .
\label{eq:main_edge_pairing_matrix_element}
\end{equation}
Here $|n,k_y\rangle$ is a BdG eigenstate of the translation-invariant ribbon, and $\Lambda_\alpha^{\rm edge}$ is the BdG coupling matrix generated by the edge pairing-fluctuation component $\alpha$. The matrix element $M_\alpha^{mn}$ measures how strongly this fluctuation scatters a quasiparticle from $k_-$ to $k_+$. Equation~\eqref{eq:main_pi_spectral} is the superconducting analog of the Lindhard response: the density matrix element is replaced by a pairing-fluctuation matrix element, and the electronic bands are replaced by BdG quasiparticle bands. Details of this calculation are given in Methods Sec.~\ref{app:edge_susceptibility}.

\begin{figure}[hb]
\centering
\includegraphics[width=1\linewidth]{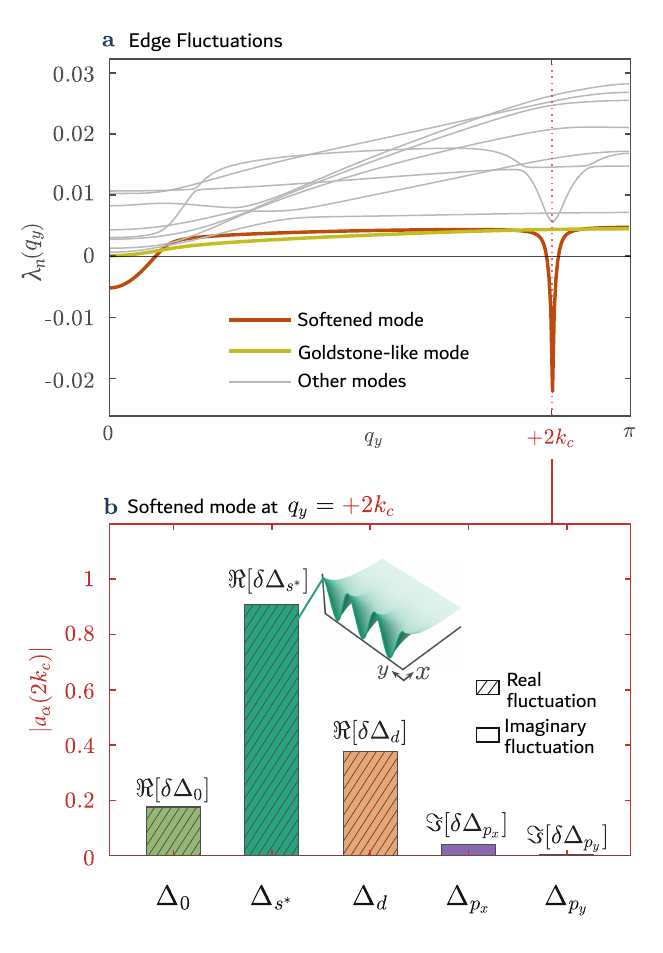}
\caption{Superconducting Kohn anomaly in the edge pairing-fluctuations. (a) Stiffness eigenvalues $\lambda_n(q_y)$ of the static edge pairing-fluctuation kernel $\mathcal K(q_y)=\mathcal K^{(0)}(q_y)-\Pi(q_y)$. The lowest (red) branch develops a cusp-like minimum at $q_y=\pm2k_c$, signalling a finite-momentum edge pairing instability. The separate negative stiffness at $q_y=0$ is the uniform ribbon-orientation instability, see text. Gold branch is the Goldstone-like branch. (b) Components of the lowest eigenvector at $q_y=+2k_c$, separated into pairing channels. Inset schematically illustrates the edge-localized, $2k_c$-modulated $\mathrm{Re}\left[\delta\Delta_{s^\ast}(x,y)\right]$ fluctuation; the other nonzero pairing-channel amplitudes correspond to similar fluctuations (not illustrated). Calculation performed for a $y$-translation-invariant ribbon with $N_x=20$ and $N_k=600$ at $U/t=-3$, $V/t=-4$, and fixed $\mu/t=-0.90493$ (see Methods Sec.~\ref{app:bdg_response_kernel}), corresponding to $n_e=0.75$.}
\label{fig:fig5}
\end{figure}

In Fig.~\ref{fig:fig5}(a) we plot the eigenvalues of the full stiffness matrix $[\mathcal K]_{\alpha\beta}(q_y)$. Before discussing the finite-momentum Kohn anomaly, we distinguish the near-zero-momentum features. The gold-coloured Goldstone-like branch near $q_y=0$ is the edge-projected remnant of the condensate phase mode, with its $q_y=0$ eigenvector corresponding to the global phase rotation $\delta\boldsymbol{\Delta}\propto i\bar{\boldsymbol{\Delta}}$, where $\bar{\boldsymbol{\Delta}}$ denotes the translation-invariant equilibrium order parameters. We refer to this as a ``Goldstone-like branch'' because the fluctuation basis is restricted to edge-localized profiles, whereas a true Goldstone mode would involve a phase rotation of the order parameters across the entire ribbon. 

Furthermore, the negative stiffness of the red branch at $q_y=0$ represents a uniform orientation instability of the chosen $s+d+ip_x$ ribbon solution toward the competing $y$-translation-invariant $s+d+ip_y$ solution, with its eigenvector dominated by $\mathrm{Im}\,\delta\Delta_{p_y}$. The finite-width ribbon lifts the bulk degeneracy between these two solutions and favours the $s+d+ip_y$ solution, as also confirmed by direct free-energy comparison. Still, fully open-boundary calculations recover the corresponding PDW on edges normal to $x$ ($y$) for the $s+d+ip_x$ ($s+d+ip_y$) orientation, confirming that the finite-wavevector boundary mechanism is independent of this ribbon orientation instability (see SM). Thus, the chosen ribbon, with edges normal to $x$, is the correct idealized system to study edge effects for the $s+d+ip_x$ state.

Next, focusing on finite momentum, we find the lowest branch (red) exhibiting a physically distinct minimum at $q_y=\pm2k_c$, precisely the wavevector connecting the two zero-energy ABS crossings in Fig.~\ref{fig:fig3}(a). This is the superconducting Kohn anomaly: the enhanced quasiparticle response at the ABS nesting wavevector softens a finite-momentum pairing mode. At the parameters considered here, this stiffness also becomes negative, signalling a finite-momentum instability of the translation-invariant state to the edge PDW. 

The softened eigenvector components at $q_y=+2k_c$ are shown in Fig.~\ref{fig:fig5}(b). This eigenvector identifies the primary instability of the translation-invariant edge at quadratic order in the pairing-fluctuations. It is multicomponent, with real fluctuations of the spin-singlet order parameters, $\mathrm{Re}\,\delta\Delta_0$, $\mathrm{Re}\,\delta\Delta_{s^\ast}$, and $\mathrm{Re}\,\delta\Delta_d$, together with an imaginary fluctuation of the $p_x$ component, $\mathrm{Im}\,\delta\Delta_{p_x}$, consistent with the dominant components of the self-consistent PDW in Fig.~\ref{fig:fig4}. 

By contrast, the $\mathrm{Im}\,\delta\Delta_{p_y}$ weight is negligible in this eigenvector, showing that $\Delta_{p_y}$ is not a primary component of the Kohn-anomaly instability.
Instead, the finite $\Delta_{p_y}$ edge modulation in the self-consistent BdG solution in Fig.~\ref{fig:fig4}(e) is induced as a secondary component by the finite-amplitude edge PDW. As shown in the SM, edge-localized Lifshitz-type gradient couplings allow $\Delta_{p_y}$ to mix with the primary PDW components, but only at finite momentum. Because these couplings involve a derivative along the edge, they convert a cosine-like primary modulation into a sine-like induced $\Delta_{p_y}$ response. This explains the $\pi/2$ spatial phase shift of $\Delta_{p_y}$ relative to the dominant $s$, $d$, and $p_x$ modulations in Fig.~\ref{fig:fig4}.

Together, Figs.~\ref{fig:fig3}--\ref{fig:fig5} establish the three-step superconducting Peierls instability sequence introduced in Sec.~\ref{sec:introduction}:
\begin{enumerate}[label=(\arabic*),leftmargin=*,itemsep=0pt,topsep=3pt]
\item The translation-invariant edge has Bogoliubov Fermi points at $\pm k_c$, formed by counter-propagating zero-energy ABS crossings [Fig.~\ref{fig:fig3}(a)].
\item The edge pairing stiffness is softened at the connecting wavevector $Q=2k_c$, producing the superconducting Kohn anomaly [Fig.~\ref{fig:fig5}(a)].
\item The softened pairing mode condenses into the self-consistent edge PDW shown in Fig.~\ref{fig:fig4}, which gaps the ABS crossings [Fig.~\ref{fig:fig3}(b)].
\end{enumerate}
The instability is therefore selected by the dispersing boundary quasiparticles, in direct analogy with the way the conventional Peierls distortion is selected by the Fermi surface.

\section{Discussion and Outlook}
\label{sec:discussion}

We identify a superconducting Peierls instability: a spontaneous, translation-symmetry-breaking reconstruction of the superconducting order parameter driven by dispersing zero-energy Bogoliubov quasiparticles. In the $s+d+ip$ example studied here, each edge of a translation-invariant ribbon hosts a dispersing ABS with zero-energy crossings at finite momenta. The edge pairing stiffness then develops a Kohn-anomaly-like softening at the wavevector connecting the counter-propagating ABS crossings. The self-consistent BdG solution consequently forms an edge-localized PDW at this wavevector, which gaps the low-energy ABS spectrum. Thus, as in the conventional Peierls mechanism, the ordering wavevector is selected by the same quasiparticles that are removed by the ordered state.

The calculations above establish the superconducting Peierls instability in the effectively zero-temperature mean-field regime, where the superconducting Kohn anomaly is sharpest. At finite temperature, thermal broadening weakens the enhanced finite-$Q$ response, suggesting a finite mean-field onset temperature for the edge PDW. Beyond mean field, low-dimensional fluctuations may further modify its finite-temperature behavior. Determining whether the mean-field onset corresponds to a sharp transition or a crossover is left for future work.

The edge PDW identified here belongs to a broader class of superconducting boundary instabilities in which spontaneous order removes low-energy Andreev or Majorana modes. Its closest comparison is to phase-crystal states at pair-breaking boundaries~\cite{Vorontsov2009BrokenSymmetryFilms,Holmvall2018BrokenTranslationEdges,Holmvall2020PhaseCrystals,WallWennerdal2020PhaseCrystal,Chakraborty2022DisorderRobustPhaseCrystal,Seja2025ImpurityPhaseCrystal,Varada2026_nss}. Both mechanisms are driven by low-energy Andreev modes. Phase crystals, however, exploit an extensive zero-energy flat band and are also primarily textures of the superconducting \emph{phase} accompanied by spontaneous currents. The superconducting Peierls mechanism instead requires only isolated crossings of a dispersive ABS and the resulting PDW is primarily an \emph{amplitude} texture. Surface PDWs in imbalanced superconductors provide another useful comparison~\cite{Barkman2019PRL_spdw}, although their finite wavevector is inherited from a bulk Fulde--Ferrell--Larkin--Ovchinnikov-type (FFLO-type) gradient instability rather than selected by boundary quasiparticles.

These comparisons isolate the defining feature of the superconducting Peierls mechanism: scattering between low-energy quasiparticles selects the ordering wavevector of the instability. Related kinematic ideas arise in translation-breaking non-superconducting states at boundaries and defects, and in Weyl-nodal systems~\cite{Kolomeisky1996sdw,Barja2016MoSe2,Wang2020MoTe2,Zhong2013PRB_cacdw,Maciejko2014PRBwswsri,Laubach2016PRB_dwi}. However, these examples all reconstruct normal electronic states through density-wave order. In our results, the reconstructed states are Bogoliubov quasiparticles, and the ordering field is the superconducting order parameter itself. The superconducting Peierls mechanism can additionally apply to boundary spectra with multiple zero-energy crossings, for which different pairs of crossings provide several candidate ordering wavevectors~\cite{Tanaka2010AnomalousABS,Daido2017MajoranaFlatBands}. It also directly indicates a possible bulk analog in gapless superconductors, where finite-$Q$ pairing-fluctuations could connect zero-energy regions of a Bogoliubov Fermi surface and thereby gap them~\cite{Agterberg2017BFS,Brydon2018BFS,Link2020NoncentroBFS}. We leave the investigation of such bulk superconducting Peierls instabilities for future work.

The criterion on the Bogoliubov quasiparticle spectrum also indicates where the superconducting Peierls mechanism might be realized. 2D or quasi-2D unconventional superconductors with well-defined edges and zero-energy boundary crossings at finite momenta are particularly promising, as illustrated with our example of the extended Hubbard model. Multicomponent and parity-mixed superconductors are promising candidates because they naturally support distinctive interface ABS~\cite{GorkovRashba2001,Frigeri2004,Smidman2017,Iniotakis2007NCSABS,Tanaka2010NCSABS}. Related systems include locally noncentrosymmetric multilayers and anapole states generated by mixed-parity interband pairing~\cite{Yoshida2014ParityMixed,Kanasugi2022Anapole}. For example, recent orientation-resolved tunnelling measurements on the noncentrosymmetric superconductor Nb$_{18}$Re$_{82}$ report a sizeable mixed-parity gap and surface spectra consistent with ABS~\cite{Strohmeier2026NbRe}, making it a potentially interesting platform for exploring the superconducting Peierls mechanism identified here. 

Experimentally, the clearest signature of the superconducting Peierls mechanism would be a boundary-localized modulation of the pair amplitude accompanied by a gap, or at least a splitting, of the low-energy ABS spectrum. Tunnelling could detect the loss or splitting of the zero-bias spectral weight, while spatially resolved probes could reveal the modulation period $2\pi/Q_{\rm PDW}$. Other boundary-induced pairing phenomena have already been discussed in several unconventional superconducting settings~\cite{Matsubara2020PRB_eodip,Takabatake2021PRB_tco,Romano2013PRL_mig}, and the observation of 1D PDW order along domain walls in monolayer Fe(Te,Se) demonstrates that such finite-momentum pairing textures can be resolved experimentally~\cite{Liu2023pdw}. Cold-atom realizations of extended Hubbard models provide a complementary route where interactions, boundaries, and edge-resolved correlations can be controlled independently~\cite{Huang2013EHM,Tarruell2018FermiHubbard}. More broadly, our results show how finite-momentum Bogoliubov quasiparticles can shape the superconducting order parameter itself, providing a new mechanism for spatially nonuniform superconducting states.

\section{Acknowledgements}
We acknowledge invaluable conversations with Andrej Mesaros, Patric Holmvall, Sushanth Varada, and other members of the Black-Schaffer Group. This work was supported in part by the Natural Sciences and Engineering Research Council of Canada (NSERC), and PSY particularly acknowledges the support of NSERC through the Canada Postdoctoral Research Award program. We acknowledge financial support from the European Union through the European Research Council (ERC) under the European Union’s Horizon 2020 research and innovation programme (ERC-2022-CoG, Grant agreement No.~101087096). Views and opinions expressed are, however, those of the authors only and do not necessarily reflect those of the European Union or the European Research Council Executive Agency. Neither the European Union nor the granting authority can be held responsible for them. The authors acknowledge the use of OpenAI's Codex and Google's Antigravity Integrated Development Environments (IDEs) and their internal AI models during the preparation of this work. These tools assisted with code development and debugging, exploratory algebraic manipulations and derivations, checking analytical expressions, and revising the presentation of the manuscript. The authors directed their use and independently verified all AI-assisted derivations, code, numerical results, references, and manuscript text. All scientific judgments and conclusions were made by the authors, who take full responsibility for the content of the manuscript.

\bibliography{biblio}

\setcounter{section}{0}
\renewcommand{\thesection}{\Alph{section}}
\renewcommand{\thesubsection}{\thesection.\arabic{subsection}}
\renewcommand{\thesubsubsection}{\thesubsection.\arabic{subsubsection}}

\makeatletter
\renewcommand{\p@subsection}{}
\renewcommand{\p@subsubsection}{}
\makeatother

\numberwithin{equation}{section}
\numberwithin{figure}{section}

\newpage
\newpage
\onecolumngrid
\begin{center}{\bf METHODS}\end{center}
\label{sec:methods}
\twocolumngrid

\section{Self-consistent BdG formalism}
\label{app:bdg_formalism}
We demonstrate the superconducting Peierls mechanism for a dominant superconducting state in the extended Hubbard model, Eq.~\eqref{eq:extended_hubbard_new}. To arrive at the necessary self-consistent Bogoliubov-de Gennes (BdG) formalism, we first decouple the interactions in the superconducting pairing channels and define the mean-field on-site order parameter
\begin{equation}
  \Delta_0(i)
  =
  U\langle
  \hat c_{i\downarrow}\hat c_{i\uparrow}
  \rangle
\end{equation}
and, for a nearest-neighbour bond $i\rightarrow i+\delta$ pairing,
\begin{equation}
  \Delta_{\delta}(i)
  =
  V\langle
  \hat c_{i+\delta,\downarrow}\hat c_{i\uparrow}
  \rangle ,
  \qquad
  \delta=\pm\hat x,\pm\hat y .
\end{equation}
We retain only these anomalous pairing mean-fields and omit self-consistent Hartree and Fock fields. A spatially uniform Hartree shift may be absorbed into the chemical potential, which is anyway adjusted to fix the density, but spatially varying density and normal bond fields are neglected. This pairing-only approximation is chosen to isolate the competition among superconducting channels and avoid normal-state charge order, bond order, or phase separation.

Because the resulting mean-field Hamiltonian contains no spin-dependent
single-particle terms and we restrict to the $S_z=0$ pairing sector, the
full spin BdG problem separates into two Nambu blocks exchanged by an
antiunitary particle--hole symmetry. We therefore diagonalize one
spin-reduced block, using the Nambu spinor
\begin{equation}
  \Psi_i
  =
  \begin{pmatrix}
    \hat c_{i\uparrow}\\
    \hat c_{i\downarrow}^\dagger
  \end{pmatrix}.
\end{equation}
This reduction still captures both spin-singlet and $S_z=0$ spin-triplet pairing, obtained from the symmetric and antisymmetric combinations of the directed bond order parameters, respectively.
In this basis, the quadratic mean-field Hamiltonian can be written as
\begin{equation}
\hham_{\rm MF}
=
\sum_{ij}
\Psi_i^\dagger
[\mathcal H_{\rm BdG}]_{ij}
\Psi_j
+
E_c,
\label{eq:app_real_space_mean_field_hamiltonian}
\end{equation}
where $E_c$ collects the fermion-independent constant energy terms generated by the mean-field decoupling and by normal ordering in the Nambu representation. Although $E_c$ does not affect the BdG eigenstates, it contributes to the mean-field free energy through its dependence on the order parameters. It must therefore be retained when comparing different self-consistent solutions, and its quadratic expansion in pairing fluctuations provides
the bare pairing kernel $\mathcal K^{(0)}(q_y)$ derived later in
Sec.~\ref{app:bare_kernel}. The corresponding real-space
BdG matrix is
\begin{equation}
\mathcal H_{\rm BdG}
=
\begin{pmatrix}
h & \Delta\\
\Delta^\dagger & -h^\ast
\end{pmatrix},
\label{eq:app_bdg_matrix}
\end{equation}
with normal-state matrix elements
\begin{equation}
  h_{ij}
  =
  -t\,\delta_{\langle ij\rangle}
  -
  \mu\,\delta_{ij}.
\end{equation}
The pairing matrix contains onsite and directed nearest-neighbour components,
\begin{equation}
  \Delta_{ij}
  =
  \Delta_0(i)\delta_{ij}
  +
  \sum_{\delta=\pm\hat x,\pm\hat y}
  \Delta_\delta(i)\delta_{j,i+\delta}.
  \label{eq:app_pairing_matrix}
\end{equation}

To reach a self-consistent solution, we solve
\begin{equation}
\mathcal H_{\rm BdG}|n\rangle=E_n|n\rangle,
\qquad
|n\rangle=
\begin{pmatrix}
u_n\\
v_n
\end{pmatrix},
\label{eq:app_bdg_eigenproblem}
\end{equation}
and update the order parameters using the electron and hole components of these eigenstates. The antiunitary particle--hole symmetry of the full BdG Hamiltonian exchanges the two spin-reduced blocks. Explicitly, if $|n\rangle=(u_n,v_n)^T$ is an eigenstate of the retained block with energy $E_n$, then $(-v_n^\ast,u_n^\ast)^T$ is an eigenstate of the complementary block with energy $-E_n$~\cite{DeGennes2018Chap5}.

The onsite order parameter is
\begin{equation}
  \Delta_0(i)
  =
  U\sum_n
  u_{in}v_{in}^\ast
  \left[
    1-2f(E_n)
  \right],
  \label{eq:app_delta0_self_consistency}
\end{equation}
while the directed bond order parameter is
\begin{equation}
\begin{aligned}
  \Delta_\delta(i)
  =
  V\sum_n
  \bigg[
    u_{in}v_{i+\delta,n}^\ast
    \left(
      1-f(E_n)
    \right)
    -
    u_{i+\delta,n}v_{in}^\ast
    f(E_n)
  \bigg].
\end{aligned}
\label{eq:app_bond_self_consistency}
\end{equation}
These equations are iterated until the order parameters converge. The electron filling is fixed through
\begin{equation}
  n_e
  =
  \frac{1}{N}
  \sum_{i,\sigma}
  \langle
  \hat n_{i\sigma}
  \rangle ,
\end{equation}
with the chemical potential adjusted to obtain the specified value of $n_e$. All calculations in the main text are performed self-consistently at the quoted filling.

\section{Superconducting order parameters}
\label{app:op_conventions}

We denote the directed bond order parameters in Eq.~\eqref{eq:app_pairing_matrix} as
\begin{equation}
\Delta_{x+},\quad
\Delta_{x-},\quad
\Delta_{y+},\quad
\Delta_{y-}.
\end{equation}
We then symmetrize them into the following pairing channels,
\begin{equation}
\begin{aligned}
\Delta_{s^\ast}
&=
\Delta_{x+}+\Delta_{x-}+\Delta_{y+}+\Delta_{y-},
\\
\Delta_d
&=
\Delta_{x+}+\Delta_{x-}-\Delta_{y+}-\Delta_{y-},
\\
\Delta_{p_x}
&=
\Delta_{x+}-\Delta_{x-},
\\
\Delta_{p_y}
&=
\Delta_{y+}-\Delta_{y-}.
\end{aligned}
\label{eq:pairing_channels}
\end{equation}
Together with the onsite component $\Delta_0$, these are the five pairing channels used throughout the work. We denote them collectively by $\Delta_\Box$, with $\Box\in\{0,s^\ast,d,p_x,p_y\}$.

For plotting the real-space order parameters in Fig.~\ref{fig:fig4}, we fix the global phase using the bulk value of $\Delta_0$ and then align each channel with its bulk phase. The plotted quantity is the real component along this phase direction. This choice displays the sign-changing PDW modulation directly, whereas plotting $|\Delta_\Box|$ would obscure the amplitude oscillations.

In the lower panels of Fig.~\ref{fig:fig4}, we further isolate the translation-breaking component at the edge by subtracting the average of each order parameter along $y$ at each fixed $x$,
\begin{equation}
\delta\Delta_\Box(x,y)
=
\mathrm{Re}\,\left[\Delta_\Box(x,y)\right]
-
\frac{1}{N_y}
\sum_{y'}
\mathrm{Re}\,\left[\Delta_\Box(x,y')\right].
\end{equation}
The edge Fourier spectrum is then computed from the boundary profile,
\begin{equation}
\begin{aligned}
\Delta_\Box(q_y)
&=
\sum_y
e^{-iq_y y}
\mathrm{Re}\,\left[\Delta_\Box(x_{\rm edge},y)\right],
\\
q_y
&=
\frac{2\pi n}{N_y}.
\end{aligned}
\end{equation}
The dominant nonzero peak of this fast Fourier transform (FFT) spectrum defines the PDW wavevector $Q_{\rm PDW}$ used in the main text.

In terms of the parent bulk state, the notation $s+d+ip$ refers to its phase structure in momentum space. For a uniform state, the pairing function may be written as 
\begin{equation}
\begin{aligned}
\Delta(\mathbf k)
=&\,
\Delta_0
+
\frac{\Delta_{s^\ast}}{2}
(\cos k_x+\cos k_y)
+
\frac{\Delta_d}{2}
(\cos k_x-\cos k_y)
\\
&+i\Delta_{p_{x,y}}\sin k_{x,y},
\end{aligned}
\label{eq:app_momentum_gap}
\end{equation}
where $p_{x,y}$ denotes either $p_x$ or $p_y$. In this convention the even-parity components $\Delta_0,\Delta_{s^\ast},\Delta_d$ are chosen real, while the odd-parity $p$-wave components carry a relative phase of $\pi/2$. This relative phase is what we denote by $s+d+ip$.
In the real-space BdG calculations, the nearest-neighbour pairing is instead represented using directed bond fields. For a uniform state, the contribution from the $x$-directed bonds is
\begin{equation}
\begin{aligned}
  \Delta_{x+}e^{ik_x}
  +
  \Delta_{x-}e^{-ik_x}
  =&\,
  (\Delta_{x+}+\Delta_{x-})\cos k_x
  \\
  &+
  i(\Delta_{x+}-\Delta_{x-})\sin k_x .
\end{aligned}
\label{eq:app_directed_bond_fourier}
\end{equation}
Thus, the Fourier transform of an antisymmetric bond combination naturally generates the factor of $i$ multiplying the odd-parity basis function in Eq.~\eqref{eq:app_momentum_gap}. The real-space channel
\begin{equation}
  \Delta_{p_x}=\Delta_{x+}-\Delta_{x-}
\end{equation}
may therefore still be real, even though it appears as $i\Delta_{p_x}\sin k_x$ in the momentum-space order parameter. The same statement applies to $\Delta_{p_y}$.

We note that the physical $\pi/2$ phase between even- and odd-parity sectors is basis independent, but the explicit location of the factor $i$ is conventional. In the figures we use the real-space bond convention natural for the BdG calculations and phase-align each order-parameter channel to its bulk value. The plotted fields therefore show the amplitude and sign of the local channel modulation, while the $s+d+ip$ phase structure is understood through the momentum-space representation in Eq.~\eqref{eq:app_momentum_gap}.

\section{Geometries and numerical setup}
\label{app:bdg_geometries}

We use a ribbon geometry in the main text and complement with fully open boundary conditions in the Supplementary Material. In the latter case, both the $x$ and $y$ directions have open boundaries and the BdG Hamiltonian is diagonalized in real space. This geometry verifies that the edge modulation develops without imposing translation symmetry along either boundary direction and thus supports the use of the ribbon geometry to isolate and characterize the boundary instability. We perform these fully open calculations for $n_e=0.75$, $U/t=-3$, and $V/t=-4$ and $-5$, using a rectangular system with $N_x=30$ and $N_y=40$.

The ribbon has open edges for $x = \{1,\,N_x\}$, see Fig.~\ref{fig:fig2}(b). When translation symmetry is imposed along $y$, $k_y$ is a good quantum number and the BdG Hamiltonian is diagonalized separately at each $N_k$ momentum. For the periodic momentum grid used here, $N_k=N_y$. The translation-invariant spectrum in Fig.~\ref{fig:fig3}(a) is calculated with $N_x=20$ and $N_k=N_y=600$. After translation symmetry along $y$ is relaxed, the BdG equations are solved fully in real space on an $N_x=20$, $N_y=150$ ribbon (with open edges at $x = \{1,\,N_x\}$ but edges at $y = \{1,\,N_y\}$ connected), allowing the edge PDW to form self-consistently. This calculation provides both the momentum-resolved PDW spectrum in Fig.~\ref{fig:fig3}(b) (see Section~\ref{app:abs_tracking} for details) and the real-space order parameters and their Fast Fourier transform (FFT) spectrum in Fig.~\ref{fig:fig4}, using $N_k=N_y=150$. For clarity, the real-space panels in Fig.~\ref{fig:fig4} display only 100 of the 150 sites along $y$.

We briefly summarize the numerical checks used to distinguish the edge PDW from possible metastable artifacts of the self-consistency procedure. The PDW solutions in the main text were obtained from several classes of initial conditions, including weakly modulated seeds, random complex bond mean-fields, and states initialized from the translation-invariant mixed-symmetry solution. When the edge PDW is stable, these initial conditions all converge to solutions with the same dominant edge wavevector and the same localization near the open boundary.

The PDW is also insensitive to changes in the transverse system size, $N_x$, once the ribbon is wide enough compared with the decay length of the edge texture, as we find that increasing $N_x$ leaves the boundary profile and Fourier peak essentially unchanged. By contrast, increasing $N_y$ sharpens the allowed momentum grid used to identify $Q_{\rm PDW}$. Within this discrete momentum resolution, the finite wavevector extracted from the self-consistent order parameters in Fig.~\ref{fig:fig4} agrees with the ABS scattering wavevector obtained independently from the translation-invariant ribbon in Fig.~\ref{fig:fig3}(a).

\section{Ribbon spectrum}
\label{app:abs_tracking}

For a ribbon open along $x$ and translation invariant along $y$, we Fourier transform the fermion operators along the periodic direction according to
\begin{equation}
\hat c_{x,k_y,\sigma}=\frac{1}{\sqrt{N_y}}\sum_y e^{-ik_y y}\hat c_{(x,y),\sigma}.
\label{eq:app_ribbon_fourier_transform}
\end{equation}
At each conserved momentum $k_y$, the spin-reduced Nambu spinor is
\begin{equation}
\Psi_{k_y}=
\begin{pmatrix}
\hat c_{1,k_y,\uparrow}\\
\vdots\\
\hat c_{N_x,k_y,\uparrow}\\
\hat c^\dagger_{1,-k_y,\downarrow}\\
\vdots\\
\hat c^\dagger_{N_x,-k_y,\downarrow}
\end{pmatrix}.
\label{eq:app_ribbon_nambu_spinor}
\end{equation}
Applying this Fourier transformation to
Eq.~\eqref{eq:app_real_space_mean_field_hamiltonian} gives
\begin{equation}
\hham_0
=
\sum_{k_y}
\Psi^\dagger_{k_y}
\mathcal H_{\rm BdG}(k_y)
\Psi_{k_y}
+
E_c,
\label{eq:app_uniform_bdg_hamiltonian}
\end{equation}
where $\hham_0$ denotes the mean-field Hamiltonian of the
translation-invariant ribbon and $E_c$ is the constant energy term. Translation invariance along $y$ decomposes the BdG problem into independent $2N_x\times2N_x$ matrices satisfying
\begin{equation}
\mathcal H_{\rm BdG}(k_y)|n,k_y\rangle
=
E_n(k_y)|n,k_y\rangle.
\label{eq:app_ribbon_bdg_eigenproblem}
\end{equation}

An eigenstate has the Nambu components
\begin{equation}
|n,k_y\rangle=
\begin{pmatrix}
u_n(1,k_y)\\
\vdots\\
u_n(N_x,k_y)\\
v_n(1,k_y)\\
\vdots\\
v_n(N_x,k_y)
\end{pmatrix},
\label{eq:app_ribbon_bdg_components}
\end{equation}
where $u_n(x,k_y)$ and $v_n(x,k_y)$ are the electron and hole amplitudes at transverse position $x$. The localization of an eigenstate across the ribbon is characterized by its center of mass,
\begin{equation}
  \bar{x}_n(k_y)
  =
  \frac{
  \sum_x x
  \left[
    |u_n(x,k_y)|^2+
    |v_n(x,k_y)|^2
  \right]
  }{
  \sum_x
  \left[
    |u_n(x,k_y)|^2+
    |v_n(x,k_y)|^2
  \right]
  } .
\end{equation}
This quantity is used to colour the spectra in Fig.~\ref{fig:fig3}: states localized on opposite edges have opposite colours, while bulk-like states lie near the center of the colour scale.

The ABS branch is identified by combining edge localization with continuity in $k_y$. Starting from a momentum where the edge state is well separated from the bulk bands, the branch is followed across the Brillouin zone by maximizing the overlap of neighbouring eigenvectors. The zero-crossing momentum is then defined by
\begin{equation}
  E_{\rm ABS}(\pm k_c)=0 ,
\end{equation}
with $k_c$ extracted by interpolating the tracked ABS dispersion between the nearest discrete momenta on either side of the sign change. 

In the fully modulated PDW state, translation symmetry along $y$ is broken and $k_y$ is no longer a good quantum number. Still, we can Fourier-resolve each real-space eigenstate along the previously periodic direction,
\begin{align}
  \tilde{u}_n(x,k_y)
  &=
  \sum_y e^{-ik_y y}u_n(x,y),
  \\
  \tilde{v}_n(x,k_y)
  &=
  \sum_y e^{-ik_y y}v_n(x,y).
\end{align}
The momentum weight plotted in Fig.~\ref{fig:fig3}(b) is therefore
\begin{equation}
  W_n(k_y)
  =
  \sum_x
  \left[
    |\tilde{u}_n(x,k_y)|^2
    +
    |\tilde{v}_n(x,k_y)|^2
  \right],
\end{equation}
normalized by the maximum weight of the same eigenstate. This unfolded spectrum shows where the PDW state carries spectral weight in the translation-invariant ribbon momentum basis. We use this procedure to compare the uniform ABS spectrum in Fig.~\ref{fig:fig3}(a) with the gapped PDW spectrum in Fig.~\ref{fig:fig3}(b).

\section{Edge susceptibility calculation}
\label{app:edge_susceptibility}
This section constructs the static edge pairing-fluctuation kernel $\mathcal K(q_y)$ used to obtain the stiffness spectrum and the eigenvector of the softened mode in Fig.~\ref{fig:fig5}. We first define the variational edge-fluctuation basis and its coupling to the BdG Hamiltonian. We then calculate the bare pairing-cost kernel $\mathcal K^{(0)}$ and the quasiparticle-response kernel $\Pi$, before combining them as $\mathcal K=\mathcal K^{(0)}-\Pi$ and diagonalizing the full stiffness kernel.

\subsection{Edge-fluctuation basis}
\label{app:edge_profile_basis}

We first decompose each pairing channel into 
\begin{equation}
\Delta_\Box(x,y)
=
\bar{\Delta}_\Box(x)
+
\eta_\Box(x,y),
\end{equation}
where $\bar{\Delta}_\Box(x)$ is the equilibrium profile of the translation-invariant ribbon and $\eta_\Box(x,y)$ is the fluctuation about this profile. After splitting each complex fluctuation into its real and imaginary parts, we denote its Fourier component along the edge by $\Phi_\alpha(x,q_y)$, where $\alpha$ labels the resulting real or imaginary fluctuation component, as in Sec.~\ref{sec:edge_susceptibility}. The main text indicates the single-profile ansatz $\Phi_\alpha(x,q_y)=a_\alpha(q_y)w(x)$ in Eq.~\eqref{eq:edge_fluctuation_ansatz}. For the numerical calculation, we generalize this ansatz to several transverse profiles,
\begin{equation}
\Phi_\alpha(x,q_y)
=
\sum_{\ell=1}^{N_{\rm prof}}
a_{\alpha\ell}(q_y)w_\ell(x),
\label{eq:app_profile_expansion}
\end{equation}
where $N_{\rm prof}$ is the number of transverse profile functions retained in the variational edge basis. For the results in Fig.~\ref{fig:fig5}, we begin with five normalized exponential seed profiles,
\begin{equation}
\widetilde w_\xi(x)={\cal N}_\xi\exp[-(x-1)/\xi],
\label{eq:app_profile_seeds}
\end{equation}
with $\xi=\{0.75,\,1.25,\,2.0,\,3.5,\,5.5\}$. Rather than selecting a single decay length, we use these seeds to span a five-dimensional transverse edge-profile subspace. We orthonormalize them and retain all five resulting basis functions $w_\ell(x)$, so that $N_{\rm prof}=5$. Diagonalization of the full kernel then determines the optimal linear combination for each fluctuation mode. 
For ease of notation, we combine the fluctuation-component and profile indices into $I=(\alpha,\ell)$ and write $a_I\equiv a_{\alpha\ell}$. Since $\alpha$ labels the real and imaginary parts of five pairing channels and $\ell=1,\ldots,5$, the fluctuation basis has dimension $50$ at each $q_y$. 

With the fluctuations expanded as in Eq.~\eqref{eq:app_profile_expansion}, the quadratic change in the free energy can be expressed as
\begin{equation}
\delta F^{(2)}
=
\frac{1}{2}
\sum_{q_y,I,J}
a_I^\ast(q_y)
[\mathcal K]_{IJ}(q_y)
a_J(q_y).
\label{eq:app_quadratic_free_energy}
\end{equation}
Following the standard Gaussian pairing-fluctuation construction~\cite{AltlandSimons2010,Protter2021PRBfia}, we write the static edge pairing-fluctuation kernel as
\begin{equation}
\mathcal K(q_y)
=
\mathcal K^{(0)}(q_y)
-
\Pi(q_y),
\label{eq:app_edge_kernel}
\end{equation}
as in Eq.~\eqref{eq:sc_peierls_new} of the main text. Here $\mathcal K^{(0)}(q_y)$ is the microscopic pairing cost projected onto the edge-fluctuation basis, while $\Pi(q_y)$ is the static BdG quasiparticle response.

Both the bare and quasiparticle contributions are constructed in the onsite and nearest-neighbour pairing-channel basis. For the nearest-neighbour channels, inverting the definitions in Eq.~\eqref{eq:pairing_channels} gives
\begin{equation}
\begin{gathered}
\delta\Delta_{+\hat{x}}
=
\frac{1}{4}
\left(
\delta\Delta_{s^\ast}
+
\delta\Delta_d
\right)
+
\frac{1}{2}\delta\Delta_{p_x},
\\
\delta\Delta_{-\hat{x}}
=
\frac{1}{4}
\left(
\delta\Delta_{s^\ast}
+
\delta\Delta_d
\right)
-
\frac{1}{2}\delta\Delta_{p_x},
\\
\delta\Delta_{+\hat{y}}
=
\frac{1}{4}
\left(
\delta\Delta_{s^\ast}
-
\delta\Delta_d
\right)
+
\frac{1}{2}\delta\Delta_{p_y},
\\
\delta\Delta_{-\hat{y}}
=
\frac{1}{4}
\left(
\delta\Delta_{s^\ast}
-
\delta\Delta_d
\right)
-
\frac{1}{2}\delta\Delta_{p_y}.
\end{gathered}
\label{eq:app_channel_to_bond}
\end{equation}

\subsection{BdG coupling matrices for finite-momentum pairing fluctuations}
\label{app:pairing_vertices}

Evaluating the quasiparticle response $\Pi(q_y)$ requires the matrix elements of the fluctuation-dependent change in the BdG Hamiltonian. Let $\mathcal V[a]$ denote the fluctuation-dependent change in the BdG matrix. Its block connecting $k_-=k_y-q_y/2$ to $k_+=k_y+q_y/2$ is
\begin{equation}
\bigl[\mathcal V[a]\bigr]_{k_+,k_-}=\sum_I a_I(q_y)\Lambda_I^{\rm edge}(k_y,q_y).
\label{eq:app_fluctuation_operator}
\end{equation}
Here $\Lambda_I^{\rm edge}$ is the profile-resolved coupling matrix associated with the basis fluctuation $a_I(q_y)$. These matrices are the profile-resolved generalizations of the coupling matrices appearing in Eq.~\eqref{eq:main_edge_pairing_matrix_element} of the main text, and their matrix elements determine the quasiparticle response through Eq.~\eqref{eq:main_pi_spectral}. Here, we construct them for each pairing channel. 

An onsite fluctuation is associated with the lattice site and simply carries the phase $e^{iq_y y_i}$. After Fourier transforming, this phase enforces $k_+-k_-=q_y$; because the two fermions occupy the same site, their form factor contains no additional $q_y$ dependence. For a bond $i\rightarrow i+\delta$, we assign the finite-$q_y$ pair-field fluctuation to the bond center,
\begin{equation}
\delta\Delta_\delta(i)
\propto
e^{iq_y(y_i+\delta_y/2)} .
\label{eq:app_bond_center_phase}
\end{equation}
The corresponding bond contribution to the pairing Hamiltonian is
\begin{equation}
\delta\hat H_\Delta=\sum_{i,\delta}\delta\Delta_\delta(i)\hat c_{i\uparrow}^\dagger\hat c_{i+\delta,\downarrow}^\dagger+{\rm H.c.}
\end{equation}
Using Eq.~\eqref{eq:app_bond_center_phase} in this expression and Fourier transforming along the periodic direction shows that the fluctuation scatters a quasiparticle from $k_-=k_y-q_y/2$ to $k_+=k_y+q_y/2$. The bond-center phase contributes the factor $e^{iq_y\delta_y/2}$, while the two fermion operators contribute the remaining relative-momentum phase. The net directed-bond form factor is therefore $e^{ik_y\delta_y}$, with no additional $q_y$ dependence. Thus the channel matrices representing this coupling depend on the center momentum $k_y=(k_++k_-)/2$.

Next, in the transverse ribbon coordinate, define $T_x$ by $(T_x)_{xx'}=\delta_{x',x+1}$. Then, define five channel matrices as
\begin{equation}
\begin{gathered}
D_0(k_y)=\mathbbm{1},
\\
D_{s^\ast}(k_y)
=
\frac{1}{4}
\left[
T_x+T_x^\dagger
+
2\cos k_y\,\mathbbm{1}
\right],
\\
D_d(k_y)
=
\frac{1}{4}
\left[
T_x+T_x^\dagger
-
2\cos k_y\,\mathbbm{1}
\right],
\\
D_{p_x}(k_y)
=
\frac{1}{2}
\left(
T_x-T_x^\dagger
\right),
\qquad
D_{p_y}(k_y)
=
i\sin k_y\,\mathbbm{1}.
\end{gathered}
\label{eq:app_symmetry_channel_vertices}
\end{equation}
The factors of $1/4$ and $1/2$ come from the inverse transformation in Eq.~\eqref{eq:app_channel_to_bond}. The $p_x$ matrix is real and antisymmetric because the $x$ direction is kept in real space; if the transverse direction were Fourier transformed, it would give the usual $i\sin k_x$ odd-parity form factor. By contrast, the $i$ in $D_{p_y}=i\sin k_y\,\mathbbm{1}$ appears explicitly because the $y$ direction has already been Fourier transformed. With the bond-center convention used here,
\begin{equation}
D_\Box(k_y,q_y)=D_\Box(k_y).
\label{eq:app_finite_q_channel_matrix}
\end{equation}

For an order-parameter channel $\Box$, a real fluctuation can now be represented by the Nambu-space coupling matrix
\begin{equation}
\Lambda_\Box^{(\mathrm{R})}(k_y,q_y)
=
\begin{pmatrix}
0 & D_\Box(k_y)\\
D_\Box^\dagger(k_y) & 0
\end{pmatrix},
\label{eq:app_real_vertex}
\end{equation}
while an imaginary fluctuation is represented by
\begin{equation}
\Lambda_\Box^{(\mathrm{I})}(k_y,q_y)
=
\begin{pmatrix}
0 & iD_\Box(k_y)\\
-iD_\Box^\dagger(k_y) & 0
\end{pmatrix}.
\label{eq:app_imag_vertex}
\end{equation}

Finally, for the fluctuation component $\alpha$, let $\Lambda_\alpha$ denote the corresponding coupling matrix from Eqs.~\eqref{eq:app_real_vertex} or \eqref{eq:app_imag_vertex}. The edge profile is attached by the symmetrized projection
\begin{equation}
\Lambda_{I}^{\rm edge}(k_y,q_y)
=
\frac{1}{2}
\left[
W_\ell\Lambda_\alpha(k_y,q_y)
+
\Lambda_\alpha(k_y,q_y)W_\ell
\right],
\label{eq:app_profile_vertex}
\end{equation}
where $W_\ell$ is the diagonal Nambu-space matrix containing $w_\ell(x)$ in both particle and hole sectors. This symmetrization evaluates the transverse profile at the center of each $x$-directed bond, using the same bond-center convention as the bare pairing-cost projection below in Sec.~\ref{app:bare_kernel}. It is these profile-resolved coupling matrices in Eq.~\eqref{eq:app_profile_vertex} that are used in Sec.~\ref{app:trace_log_response} to evaluate the quasiparticle-response kernel $\Pi(q_y)$. The bare pairing-cost contribution is constructed separately in the following subsection using the same edge-fluctuation basis.

\subsection{Bare pairing kernel}
\label{app:bare_kernel}

We next obtain the bare kernel $\mathcal K^{(0)}(q_y)$ by expanding the $E_c$ term in Eq.~\eqref{eq:app_real_space_mean_field_hamiltonian} to quadratic order in the fluctuations and projecting the resulting microscopic pairing cost onto the edge-fluctuation basis. In the normalization appropriate to one spin-reduced Nambu block, the pairing cost is
\begin{equation}
F_{\rm pair}=-\sum_i\frac{|\Delta_0(i)|^2}{2U}-\sum_{i,\delta}\frac{|\Delta_\delta(i)|^2}{2V},
\label{eq:app_pairing_cost}
\end{equation}
with $\delta=\pm\hat{x},\pm\hat{y}$ labelling the directed nearest-neighbour fields. Note that the physical free energy quoted in the Supplementary Material includes both redundant spin blocks and is thus twice this reduced functional, but the overall factor does not affect the signs, eigenvectors, or momentum dependence of the stiffness kernel. The Hessian therefore carries the positive coefficients $-1/U$ for onsite fields and $-1/V$ for bond fields.

Using the directed-bond relations in Eq.~\eqref{eq:app_channel_to_bond}, we project this pairing cost into the transverse profile basis. This gives two overlap matrices. For onsite and edge-parallel, $y$-directed bonds,
\begin{equation}
Y_{\ell\ell'}
=
\sum_x w_\ell(x)w_{\ell'}(x).
\end{equation}
Because the transverse basis functions are orthonormal, $Y_{\ell\ell'}=\delta_{\ell\ell'}$. For transverse, $x$-directed bonds, the overlap is instead evaluated at the bond center,
\begin{equation}
X_{\ell\ell'}
=
\sum_{x=1}^{N_x-1}
\frac{w_\ell(x)+w_\ell(x+1)}{2}
\frac{w_{\ell'}(x)+w_{\ell'}(x+1)}{2}.
\end{equation}

In terms of these overlap matrices, the onsite block of the bare kernel is
\begin{equation}
[\mathcal K^{(0)}]_{00,\ell\ell'}
=
-Y_{\ell\ell'}/U ,
\label{eq:app_bare_metric_onsite}
\end{equation}
while its nearest-neighbour channel block is
\begin{equation}
[\mathcal K^{(0)}]_{rr',\ell\ell'}
=
-\frac{1}{V}
\sum_{\delta=\pm\hat{x},\pm\hat{y}}
s_{\delta r}s_{\delta r'}
Z_{\delta,\ell\ell'},
\label{eq:app_bare_metric_bond}
\end{equation}
where $r,r'\in\{s^\ast,d,p_x,p_y\}$, $Z_{\pm\hat{x},\ell\ell'}=X_{\ell\ell'}$ and $Z_{\pm\hat{y},\ell\ell'}=Y_{\ell\ell'}$. The coefficients $s_{\delta r}$ are extracted from Eq.~\eqref{eq:app_channel_to_bond}; for example $s_{+\hat{x},p_x}=1/2$, $s_{-\hat{x},p_x}=-1/2$, and $s_{\pm\hat{y},p_x}=0$. Together, Eqs.~\eqref{eq:app_bare_metric_onsite} and \eqref{eq:app_bare_metric_bond} give the bare contribution $\mathcal K^{(0)}(q_y)$: an onsite block and a four-channel nearest-neighbour block, repeated identically in the real and imaginary fluctuation sectors.

\subsection{Quasiparticle-response kernel}
\label{app:trace_log_response}

We now derive the spectral expression for the static quasiparticle-response kernel $\Pi(q_y)$ given in Eq.~\eqref{eq:main_pi_spectral} of the main text. This contribution is obtained by expanding the BdG quasiparticle free energy to quadratic order in the finite-momentum pairing fluctuations, using the coupling matrices constructed in Sec.~\ref{app:pairing_vertices}. We start with the equilibrium Nambu Green's function of the translation-invariant ribbon,
\begin{equation}
  G_0(k_y,i\omega_n)
  =
  \left[
    i\omega_n-\mathcal H_{\rm BdG}(k_y)
  \right]^{-1},
  \quad
  \omega_n=(2n+1)\pi T .
  \label{eq:app_g0}
\end{equation}
The fluctuation-dependent BdG term $\mathcal V[a]$ was constructed in Eq.~\eqref{eq:app_fluctuation_operator}. A component with momentum $q_y$ connects a state at $k_-=k_y-q_y/2$ to one at $k_+=k_y+q_y/2$. For real fluctuation fields, $a_I(-q_y)=a_I^\ast(q_y)$, so the $-q_y$ component gives the Hermitian-conjugate process. The inverse Green's function in the presence of these pairing fluctuations is therefore
\begin{equation}
G^{-1}[a]=G_0^{-1}-\mathcal V[a].
\label{eq:app_inverse_green}
\end{equation}
After integrating out the BdG quasiparticles, the static quasiparticle contribution to the free energy is, up to terms independent of $a_I$,
\begin{equation}
F_{\rm qp}[a]=-\frac{T}{2}{\rm Tr}\ln[-G^{-1}[a]].
\label{eq:app_fermion_determinant}
\end{equation}
The prefactor $1/2$ is the standard Nambu prefactor: the trace is over the full particle--hole BdG spectrum and would otherwise double count physical processes. Factoring $G^{-1}[a]=G_0^{-1}[1-G_0\mathcal V[a]]$ and expanding to second order gives
\begin{align}
F_{\rm qp}[a]&={\rm const.}-\frac{T}{2}{\rm Tr}\ln[1-G_0\mathcal V[a]]\notag\\
&={\rm const.}+\frac{T}{2}{\rm Tr}(G_0\mathcal V)+\frac{T}{4}{\rm Tr}(G_0\mathcal V G_0\mathcal V)+O(\mathcal V^3).
\label{eq:app_trace_log_expansion}
\end{align}
For $q_y\neq0$, the linear term in Eq.~\eqref{eq:app_trace_log_expansion} vanishes by momentum conservation in the translation-invariant ribbon. At $q_y=0$, the quasiparticle linear term need not vanish separately; instead, the stationarity of the self-consistent BdG solution ensures that it cancels the corresponding linear term from the pairing cost. The total thermodynamic functional therefore has no linear term. 

Matching the quadratic term in Eq.~\eqref{eq:app_trace_log_expansion} to the convention
\begin{equation}
  \delta F_{\rm qp}^{(2)}
  =
  -
  \frac{1}{2}
  \sum_{q_y,I,J}
  a_I^\ast(q_y)
  \Pi_{IJ}(q_y)
  a_J(q_y),
  \label{eq:app_qp_quadratic}
\end{equation}
gives
\begin{equation}
\begin{aligned}
\Pi_{IJ}(q_y)
&=
-\frac{T}{2N_y}
\sum_{k_y,\omega_n}
{\rm Tr}
\big[
G_0(k_+,i\omega_n)
\Lambda_J^{\rm edge}(k_y,q_y)
\\
&\qquad\qquad\qquad\quad\times
G_0(k_-,i\omega_n)
[\Lambda_I^{\rm edge}(k_y,q_y)]^\dagger
\big].
\end{aligned}
\label{eq:app_green_bubble_pi}
\end{equation}

\begin{figure}[h]
\centering
\begin{tikzpicture}[scale=0.95,fermion/.style={very thick,postaction={decorate},decoration={markings,mark=at position 0.55 with {\arrow{Latex}}}},external/.style={thick,dashed,-{Latex}}]
\coordinate (L) at (-1.15,0);
\coordinate (R) at (1.15,0);
\draw[external] (-2.15,0) -- (L) node[midway,above=2pt] {$a_J(q_y)$};
\draw[external] (R) -- (2.15,0) node[midway,above=2pt] {$a_I^\ast(q_y)$};
\draw[fermion] (L) .. controls (-0.75,0.85) and (0.75,0.85) .. node[midway,above=5pt] {$G_0(k_+,i\omega_n)$} (R);
\draw[fermion] (R) .. controls (0.75,-0.85) and (-0.75,-0.85) .. node[midway,below=5pt] {$G_0(k_-,i\omega_n)$} (L);
\fill (L) circle (1.7pt) node[below left=3pt] {$\Lambda_J^{\rm edge}$};
\fill (R) circle (1.7pt) node[below right=3pt] {$[\Lambda_I^{\rm edge}]^\dagger$};
\end{tikzpicture}
\caption{Feynman diagram for the quasiparticle-response bubble in Eq.~\eqref{eq:app_green_bubble_pi}. The external pairing fluctuation transfers momentum $q_y$, so the two BdG propagators carry $k_\pm=k_y\pm q_y/2$ at the same fermionic Matsubara frequency $i\omega_n$.}
\label{fig:app_pairing_bubble}
\end{figure}
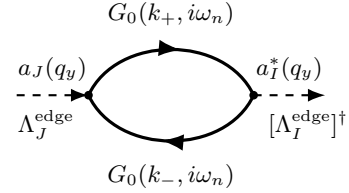

The trace is over the $2N_x$ Nambu degrees of freedom of the ribbon, and the prefactor $1/N_y$ follows from the Fourier normalization along the periodic direction. The corresponding quasiparticle bubble is shown diagrammatically in Fig.~\ref{fig:app_pairing_bubble}. 

To obtain the spectral form, we insert the eigenstate expansion of the unperturbed BdG Green's function,
\begin{equation}
  G_0(k_y,z)
  =
  \sum_n
  \frac{
    |n,k_y\rangle\langle n,k_y|
  }{
    z-E_n(k_y)
  } .
  \label{eq:app_green_spectral_expansion}
\end{equation}
We then define the pairing-fluctuation matrix element
\begin{equation}
  M_I^{mn}(k_y,q_y)
  =
  \langle m,k_+|
  \Lambda_I^{\rm edge}(k_y,q_y)
  |n,k_-\rangle ,
  \label{eq:app_pairing_matrix_element_from_trace}
\end{equation}
which gives
\begin{equation}
\begin{aligned}
  \Pi_{IJ}(q_y)
  &=
  -\frac{1}{2N_y}
  \sum_{k_y,m,n}
  M_J^{mn}(k_y,q_y)\left[M_I^{mn}(k_y,q_y)\right]^\ast
  \\
  &\quad\times
  T\sum_{\omega_n}
  \frac{1}{
    [i\omega_n-E_m(k_+)]
    [i\omega_n-E_n(k_-)]
  } .
\end{aligned}
  \label{eq:app_before_matsubara_sum}
\end{equation}
Evaluating the Matsubara sum as
\begin{equation}
\begin{aligned}
&T\sum_{\omega_n}
\frac{1}{
[i\omega_n-E_m(k_+)]
[i\omega_n-E_n(k_-)]
}
\\
&\qquad =
\frac{
f(E_m(k_+))-f(E_n(k_-))
}{
E_m(k_+)-E_n(k_-)
},
\end{aligned}
\label{eq:app_matsubara_sum}
\end{equation}
gives
\begin{equation}
\begin{aligned}
  \Pi_{IJ}(q_y)
  &=
  \frac{1}{2N_y}
  \sum_{k_y,m,n}
  M_J^{mn}(k_y,q_y)\left[M_I^{mn}(k_y,q_y)\right]^\ast
  \\
  &\quad\times
  \frac{
    f(E_n(k_-))-f(E_m(k_+))
  }{
    E_m(k_+)-E_n(k_-)
  } .
\end{aligned}
  \label{eq:app_pi_spectral}
\end{equation}
This is Eq.~\eqref{eq:main_pi_spectral} in the main text.
When $E_m(k_+)=E_n(k_-)=E$, the quotient in Eq.~\eqref{eq:app_pi_spectral} is understood as its limiting value $-f'(E)\geq0$. The quotient is therefore non-negative for every pair of states, making $\Pi(q_y)$ positive semidefinite. Because it enters $\mathcal K(q_y)=\mathcal K^{(0)}(q_y)-\Pi(q_y)$ with a minus sign, the quasiparticle response thus always softens the pairing stiffness.

For the results in Fig.~\ref{fig:fig5}, the profile-resolved response in Eq.~\eqref{eq:app_pi_spectral} is evaluated using the full particle--hole BdG spectrum of the translation-invariant ribbon. This provides the quasiparticle contribution $\Pi(q_y)$ to the full stiffness kernel.

\subsection{Full stiffness kernel and eigenmodes}
\label{app:bdg_response_kernel}

The full stiffness kernel used to produce Fig.~\ref{fig:fig5} is $\mathcal K(q_y)=\mathcal K^{(0)}(q_y)-\Pi(q_y)$, as defined in Eq.~\eqref{eq:app_edge_kernel}. The bare contribution is given by Eqs.~\eqref{eq:app_bare_metric_onsite} and \eqref{eq:app_bare_metric_bond}, while the quasiparticle response is given by Eq.~\eqref{eq:app_pi_spectral}. We then obtain the stiffness eigenvalues and eigenmodes from
\begin{equation}
\sum_J\mathcal K_{IJ}(q_y)b_J^{(n)}(q_y)=\lambda_n(q_y)b_I^{(n)}(q_y),
\label{eq:app_edge_kernel_eigenvalues}
\end{equation}
where $I$ and $J$ are the composite fluctuation-component/profile indices defined above Eq.~\eqref{eq:app_quadratic_free_energy}, and $b_I^{(n)}(q_y)$ is the normalized amplitude of basis fluctuation $I$ in eigenmode $n$. For each $q_y$, $\mathcal K(q_y)$ is Hermitian, so its eigenvalues are real. A negative eigenvalue identifies a fluctuation that lowers the free energy at quadratic order and therefore signals an instability of the translation-invariant solution. The stiffness eigenvalues $\lambda_n(q_y)$ are plotted in Fig.~\ref{fig:fig5}(a). The kernel is evaluated at fixed chemical potential and therefore represents the curvature of the mean-field grand potential. At fixed particle density, the chemical potential may acquire a compensating shift for uniform, $q_y=0$, fluctuations, thereby modifying the strictly uniform sector of the kernel. However, this does not affect the finite-momentum anomaly at $q_y=2k_c$ considered here.

To reconstruct the softened eigenmode in real space, we unpack the composite index as $I=(\alpha,\ell)$ and denote the components of the lowest-stiffness eigenmode at $q_y=2k_c$ by $b_{\alpha\ell}^{(\min)}(2k_c)$. Limiting the profile expansion in Eq.~\eqref{eq:app_profile_expansion} to this eigenmode gives
\begin{equation}
\Phi_\alpha^{(\min)}(x,2k_c)=\sum_{\ell=1}^{N_{\rm prof}}b_{\alpha\ell}^{(\min)}(2k_c)w_\ell(x).
\label{eq:app_soft_mode_reconstruction}
\end{equation}
In Fig.~\ref{fig:fig5}(b), we plot the total weight of each reconstructed pairing component
\begin{equation}
|a_\alpha(2k_c)|\equiv\left[\sum_x\left|\Phi_\alpha^{(\min)}(x,2k_c)\right|^2\right]^{1/2}.
\label{eq:app_channel_weight}
\end{equation}
Here $\alpha$ labels a real or imaginary pairing component. The coefficients $b_{\alpha\ell}^{(\min)}$ give the contributions of the individual transverse profiles to the normalized stiffness eigenmode, whereas $|a_\alpha|$ is the total weight of component $\alpha$ after reconstructing and combining all five profiles. The reconstructed softened mode decays over approximately two lattice spacings, much less than the ribbon width $N_x=20$, confirming that it is indeed localized to a single edge.

The real and imaginary labels in Fig.~\ref{fig:fig5}(b) follow the momentum-space gap convention of Sec.~\ref{app:op_conventions}. For the even-parity channels, they coincide with the real and imaginary parts of the corresponding onsite or directed-bond amplitudes. For the odd-parity channels, a real antisymmetric $x$-bond fluctuation contributes
\begin{equation}
\delta\Delta_{p_x}\left(e^{ik_x}-e^{-ik_x}\right)/2=i\delta\Delta_{p_x}\sin k_x
\end{equation}
to the momentum-space gap function, and similarly for $p_y$. A real coefficient of an antisymmetric $p$-wave bond fluctuation is therefore displayed as an imaginary $p$-wave gap fluctuation, consistent with the convention used to describe the bulk state as $s+d+ip_x$.
\clearpage
\onecolumngrid

\setcounter{section}{0}
\renewcommand{\thesection}{\Roman{section}}
\renewcommand{\thesubsection}{\Alph{subsection}}

\makeatletter
\@removefromreset{equation}{section}
\@removefromreset{figure}{section}
\makeatother

\setcounter{equation}{0}
\renewcommand{\theequation}{\arabic{equation}}

\setcounter{figure}{0}
\renewcommand{\thefigure}{\arabic{figure}}

\setcounter{table}{0}
\renewcommand{\thetable}{\Roman{table}}

\renewcommand{\theHsection}{supp.\Roman{section}}
\renewcommand{\theHsubsection}{supp.\Roman{section}.\Alph{subsection}}
\renewcommand{\theHequation}{supp.\arabic{equation}}
\renewcommand{\theHfigure}{supp.\arabic{figure}}
\renewcommand{\theHtable}{supp.\arabic{table}}

\addtocontents{toc}{\protect\ShowInSupplementTOC}

\title{Supplementary Material for ``A Superconducting Peierls Instability''}

\makeatletter
\let\frontmatter@abstract@produce\relax
\makeatother

\repeatablemaketitle
\onecolumngrid

\noindent The Supplementary Material (SM) is organized as follows. Section~\ref{app:abs_analytic_estimate} derives an analytic estimate of the Andreev bound state (ABS) dispersion and zero-crossing momentum $k_c$ used to interpret Fig.~3 of the main text. Section~\ref{app:abs_genericity} maps the dependence of $k_c$ on the interactions $U$ and $V$, showing that the required zero-energy crossings occur across a broad region of the $s+d+ip$ phase diagram. Section~\ref{supp:counter-propagating_kinematics} explains why a finite wavevector connecting counter-propagating zero-energy Bogoliubov states (i.e.\ states with opposite group velocities) produces the enhanced response required for the Peierls mechanism, whereas connecting co-propagating states (i.e.\ states with group velocities of the same sign) does not. Section~\ref{supp:interaction_dependence} then shows how the superconducting Kohn anomaly in Fig.~5(a) of the main text evolves with the nearest-neighbour interaction $V$ and distinguishes finite-wavevector softening from the onset of an edge pair-density wave (PDW) instability. Section~\ref{supp:free_energy_comparison} introduces the mean-field free-energy expression used to compare the self-consistent Bogoliubov-de Gennes (BdG) calculations and presents the solutions for fully open-boundary and ribbon geometries. These calculations establish the energetic favourability of the edge PDW within the selected $s+d+ip_x$ sector and distinguish the finite-wavevector Peierls instability from the separate $q_y=0$ ribbon-orientation instability. Section~\ref{app:symmetry_structure} explains the symmetry origin of the bulk $d\rightarrow s+d+ip$ transition and the multicomponent structure of the edge PDW. Finally, Section~\ref{app:abs_projection_pair_field} relates the gap-opening scattering between the zero-energy ABS states to finite-momentum Cooper-pair amplitudes with $Q=\pm2k_c$ in the microscopic electron basis.

\tableofcontents

\section{Analytic ABS Dispersion}
\label{app:abs_analytic_estimate}

This section gives the derivation of the analytical ABS dispersion shown by the dashed curves in Fig.~3(a) of the main text. For an edge normal to $x$, the conserved momentum is $k_y$, and specular reflection sends $k_x\rightarrow-k_x$. Near the boundary of the translation-invariant ribbon, we approximate the gap function by
\begin{equation}
  \Delta(\mathbf{k})=\tilde{\Delta}_{0}+\frac{\tilde{\Delta}_{s^\ast}}{2}(\cos k_x+\cos k_y)+\frac{\tilde{\Delta}_{d}}{2}(\cos k_x-\cos k_y)+i\tilde{\Delta}_{p_x}\sin k_x .
  \label{eq:app_gap_function_abs}
\end{equation}
The tildes denote representative edge values extracted from the self-consistent, $x$-dependent order-parameter profiles. Equation~\eqref{eq:app_gap_function_abs} therefore provides a local approximation that neglects their residual variation over the localization length of the ABS. At fixed $k_y$, the normal-state Fermi surface determines $k_{Fx}$ through
\begin{equation}
  -2t(\cos k_{Fx}+\cos k_y)-\mu=0 .
\end{equation}
The incoming and reflected trajectories therefore experience
\begin{equation}
  \Delta_\pm(k_y)=A(k_y)\pm iB(k_y),
  \label{eq:app_delta_pm}
\end{equation}
where
\begin{equation}
  A(k_y)=\tilde{\Delta}_{0}-\frac{\mu}{4t}\left(\tilde{\Delta}_{s^\ast}+\tilde{\Delta}_{d}\right)-\tilde{\Delta}_{d}\cos k_y,\qquad B(k_y)=\tilde{\Delta}_{p_x}\sin k_{Fx}.
  \label{eq:app_A_B_ky}
\end{equation}
In Eq.~\eqref{eq:app_gap_function_abs}, we have chosen the global gauge in which the even-parity components are real and the $p_x$ contribution to the gap is purely imaginary, so that $A(k_y)$ and $B(k_y)$ are real. Defining $\theta(k_y)$ as the complex phase of $A(k_y)+iB(k_y)$, we may write $\Delta_\pm(k_y)=|\Delta(k_y)|e^{\pm i\theta(k_y)}$. The incoming and reflected trajectories therefore experience a relative superconducting phase difference of $2\theta(k_y)$. In this sense, an edge normal to $x$ is pair-breaking for the $s+d+ip_x$ state: specular reflection reverses the reflection-odd $p_x$ component and produces the phase mismatch that supports a localized ABS. For the symmetry-related $s+d+ip_y$ state, the same pair-breaking condition instead occurs at edges normal to $y$. The reflection-induced phase difference determines the condition for forming an ABS at the edge. For specular reflection, the ABS quantization condition follows from the total phase accumulated around a closed quasiparticle--hole trajectory that samples the two pair potentials $\Delta_\pm(k_y)$ defined in Eq.~\eqref{eq:app_delta_pm}~\cite{KashiwayaTanaka2000,Lofwander2001ABS}. Because $\Delta_\pm=A\pm iB$, the two trajectories experience equal gap magnitudes, $|\Delta_+|=|\Delta_-|\equiv|\Delta|=\sqrt{A^2+B^2}$. The quantization condition for one boundary orientation can therefore be written as
\begin{equation}
2\arccos\!\left[\frac{E_{\rm ABS}(k_y)}{|\Delta(k_y)|}\right]-2\theta(k_y)=0 .
\end{equation}
This condition fixes the bound-state energy for a chosen boundary orientation; reversing the boundary orientation gives the opposite-energy branch. The two ribbon edges therefore have the dispersions
\begin{equation}
E_{\rm ABS}(k_y)=\pm|\Delta(k_y)|\cos\theta(k_y)=\pm A(k_y)=\pm\left[\tilde{\Delta}_0-\frac{\mu}{4t}\left(\tilde{\Delta}_{s^\ast}+\tilde{\Delta}_d\right)-\tilde{\Delta}_d\cos k_y\right],
\label{eq:app_abs_dispersion}
\end{equation}
where the $\pm$ solutions correspond to the two opposite edges. Thus the $p_x$ component does not appear explicitly in the ABS energy, but it produces the reflected phase difference and controls the localization of the bound state through $B(k_y)$. Equation~\eqref{eq:app_abs_dispersion} is plotted in Fig.~3(a) of the main text using numerically extracted order parameters values at the edge.

A localized zero-energy crossing therefore requires both $A(k_c)=0$ and $B(k_c)\neq0$. The analytic zero-energy crossing condition for the ABS is then given by
\begin{equation}
  \cos k_c=\frac{\tilde{\Delta}_{0}-\frac{\mu}{4t}\left(\tilde{\Delta}_{s^\ast}+\tilde{\Delta}_{d}\right)}{\tilde{\Delta}_{d}} .
  \label{eq:app_kc_estimate}
\end{equation}
For Eq.~\eqref{eq:app_kc_estimate} to describe a physical ABS crossing, its right-hand side must yield a real $k_c$ within the projected normal-state Fermi surface, and the reflection-odd component must remain nonzero, $B(k_c)\neq0$. Equation~\eqref{eq:app_kc_estimate} is an analytic estimate used to expose the origin and parameter dependence of the zero-energy crossing; it is not used to generate the numerical results. All values of $k_c$ shown in Fig.~\ref{fig:supp_abs_crossing_map} and reported in the main text are instead obtained by locating the zero-energy crossings directly in the self-consistent $y$-translation-invariant ribbon BdG spectrum.

\section{Interaction Dependence of the ABS Zero-Energy Crossings}
\label{app:abs_genericity}

\begin{figure}[ht]
\centering
\includegraphics[width=0.5\linewidth]{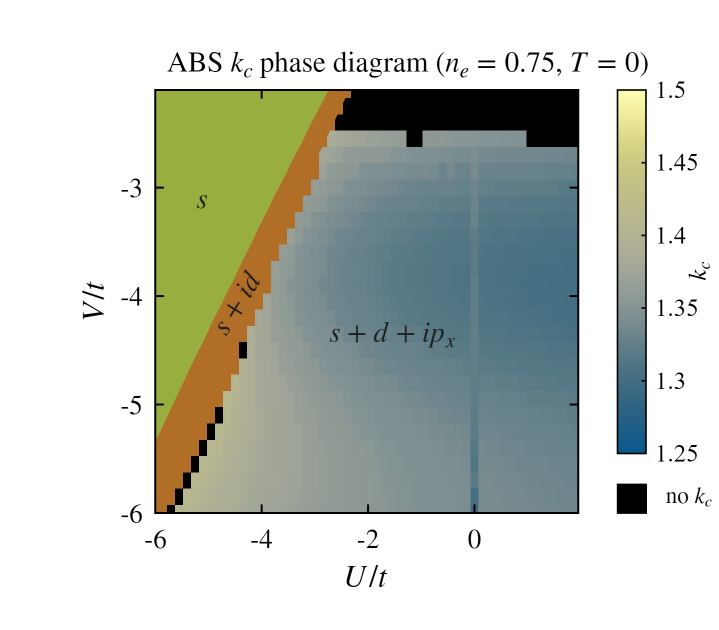}
\caption{$U$--$V$ dependence of the ABS zero-crossing momentum $k_c$ at $n_e=0.75$, extracted from translation-invariant ribbon calculations with $N_x=20$ and $N_k=400$ sampled edge momenta. The $s$ and $s+id$ regions are identified as in Fig.~2 of the main text and contain no zero-energy ABS crossing for this edge orientation. Within the $s+d+ip_x$ region, the colour scale gives $k_c$ wherever a zero-energy crossing is found, while black indicates regions where no such crossing is identified. The ribbon phase boundaries differ slightly from those of the bulk phase diagram in Fig.~2 because the self-consistent order parameters become inhomogeneous near the open boundaries, slightly modifying the relative free energies of the competing solutions.}
\label{fig:supp_abs_crossing_map}
\end{figure}

The main text demonstrates the superconducting Peierls mechanism at the representative point $n_e=0.75$, $U/t=-3$, and $V/t=-4$ in the extended Hubbard model. We now determine how broadly the required pair of finite-momentum zero-energy ABS crossings occurs as the interactions are varied. At each point in the $U$--$V$ plane, we therefore solve the translation-invariant ribbon self-consistently and numerically extract $k_c$ directly from the zero-energy crossings of its BdG spectrum. The analytic estimate in Eq.~\eqref{eq:app_kc_estimate} serves only to explain why the extracted $k_c$ varies with the chemical potential and the boundary order parameters. Figure~\ref{fig:supp_abs_crossing_map} plots the resulting map for the $s+d+ip_x$ ribbon state and shows that the required ABS crossings occur across a broad part of the $s+d+ip_x$ region. Within this region, $k_c$ varies smoothly with the microscopic parameters, making the PDW wavelength selected by $Q_{\rm PDW}=2k_c$ tunable through the interactions.

Importantly, Fig.~\ref{fig:supp_abs_crossing_map} is not a phase diagram of the edge PDW. A finite $k_c$ identifies a candidate scattering wavevector for the boundary ABS but does not by itself establish an instability of the translation-invariant edge. In addition, the quasiparticle response must soften the edge pairing stiffness at $q_y=2k_c$ sufficiently to drive an eigenvalue of the full stiffness kernel $\mathcal K(q_y)$ negative; the interaction dependence of this criterion is examined in Sec.~\ref{supp:interaction_dependence}. Mapping where the resulting finite-amplitude PDW is energetically favoured would require self-consistent solutions and free-energy comparisons throughout the parameter space. A complete phase diagram of the edge PDW is therefore left for future work.

\section{Counter-propagating and Co-propagating Bogoliubov-State Kinematics}
\label{supp:counter-propagating_kinematics}

The main text applies the superconducting Peierls mechanism to two zero-energy ABS states at distinct edge momenta. The kinematic argument, however, is more general and requires only a finite wavevector connecting zero-energy Bogoliubov states with opposite group velocities. These states may arise from boundary ABS or, more generally, from zero-energy bulk Bogoliubov bands. For the boundary realization, the two states must additionally be localized on the same physical edge so that the relevant pairing-fluctuation matrix element is nonzero; this spatial-overlap condition is separate from the kinematic requirement developed below. 

\begin{figure*}[t]
\centering
\includegraphics[width=0.9\textwidth]{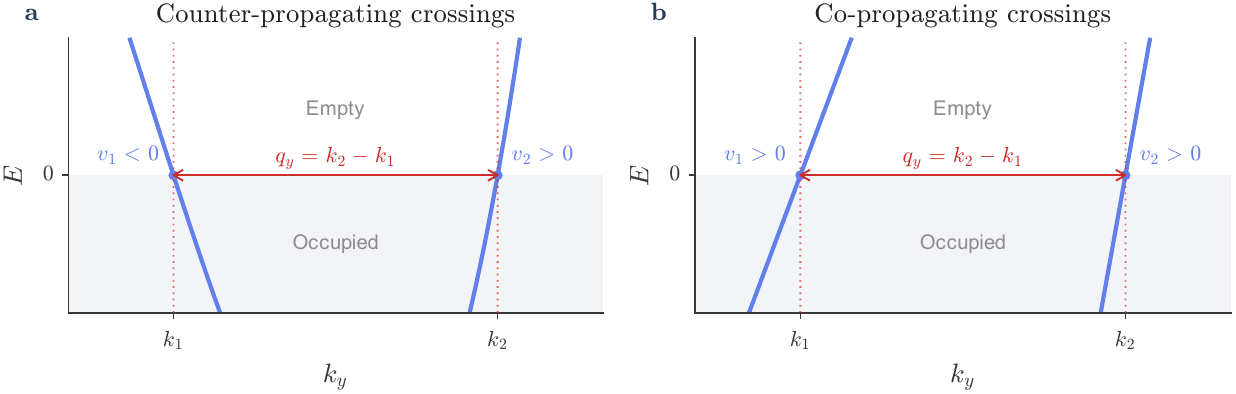}
\caption{Low-energy kinematics near two zero-energy crossings of Bogoliubov bands at distinct momenta. The illustrated states may belong to boundary ABS or to bulk Bogoliubov bands. Dotted lines mark the crossing momenta $k_1$ and $k_2$, and the red arrow shows the connecting momentum $q_y=k_2-k_1$. States below (above) zero energy are occupied (empty) at zero temperature. (a) For counter-propagating crossings, the group velocities have opposite signs. (b) For chiral or co-propagating crossings, the velocities have the same sign.
}
\label{fig:supp_bogoliubov_kinematics}
\end{figure*}

The requirement of opposite group velocities follows directly from the spectral representation of the quasiparticle response given in Eq.~(18) of the main text, 
\begin{equation}
\Pi_{\alpha\beta}(q_y)
=
\frac{1}{2N_y}
\sum_{k_y,m,n}
M_\beta^{mn}
\left[M_\alpha^{mn}\right]^\ast
\frac{
f(E_n(k_-))-f(E_m(k_+))
}{
E_m(k_+)-E_n(k_-)
},
\label{eq:supp_pi_spectral_reproduced}
\end{equation}
where $k_\pm=k_y\pm q_y/2$ and $M_\alpha^{mn}$ is the matrix element of the pairing fluctuation between the corresponding BdG eigenstates. The possible low-energy enhancement is controlled by the occupation-over-energy-denominator factor in Eq.~\eqref{eq:supp_pi_spectral_reproduced}. To see how this affects the quasiparticle response, consider two zero-energy states at momenta $k_1$ and $k_2$, connected by $q_y=k_2-k_1$, as illustrated in Fig.~\ref{fig:supp_bogoliubov_kinematics}(a,b) for counter-propagating and chiral/co-propagating branches, respectively. We compare states displaced by the same small momentum $\delta k$ from these momenta and denote their energies by $E_1(\delta k)$ and $E_2(\delta k)$, respectively. These are the two energies entering Eq.~\eqref{eq:supp_pi_spectral_reproduced}, with $k_-=k_1+\delta k$ and $k_+=k_2+\delta k$. For the ribbon studied in the main text, $k_1=-k_c$ and $k_2=+k_c$, giving $q_y=2k_c$.
To compare the two cases directly, we define the factor
\begin{equation}
\mathcal Q(\delta k)
\equiv
\frac{
f(E_1(\delta k))-f(E_2(\delta k))
}{
E_2(\delta k)-E_1(\delta k)
}.
\label{eq:app_kinematic_quotient}
\end{equation}
This is the occupation-over-energy-denominator factor in Eq.~\eqref{eq:supp_pi_spectral_reproduced} for scattering between the two crossings. For the counter-propagating crossings in Fig.~\ref{fig:supp_bogoliubov_kinematics}(a), the linearized dispersions are
\begin{equation}
E_1(\delta k)\simeq v_1\delta k,
\qquad
E_2(\delta k)\simeq v_2\delta k,
\qquad
v_1<0<v_2,
\label{eq:app_counter_propagating_linearization}
\end{equation}
where the velocity magnitudes need not be equal, $|v_1|\neq|v_2|$. At zero temperature, $f(E)=\Theta(-E)$, and Eq.~\eqref{eq:app_kinematic_quotient} becomes
\begin{equation}
\begin{aligned}
\mathcal Q_{\rm ctr}(\delta k)
&\simeq
\frac{
\Theta(-v_1\delta k)-\Theta(-v_2\delta k)
}{
(v_2-v_1)\delta k
}\\
&=
\frac{\operatorname{sgn}(\delta k)}
{(v_2-v_1)\delta k}
\\
&=
\frac{1}
{(v_2-v_1)|\delta k|},
\qquad
(\delta k\neq0).
\end{aligned}
\label{eq:app_counter_propagating_quotient}
\end{equation}
As shown in Fig.~\ref{fig:supp_bogoliubov_kinematics}(a), the states connected at a common $\delta k$ have opposite energies and hence different occupations. The numerator of Eq.~\eqref{eq:app_counter_propagating_quotient} therefore remains nonzero while the energy denominator vanishes linearly. Provided the corresponding pairing-fluctuation matrix element is nonzero, the momentum sum for the one-dimensional (1D) edge considered here then produces the logarithmic Peierls enhancement. The counter-propagating kinematic criterion is more general, but the resulting phase-space enhancement depends on dimensionality. For the reflection-symmetric ribbon in the main text, $v_1=-v$ and $v_2=v$, so Eq.~\eqref{eq:app_counter_propagating_quotient} reduces to $1/(2v|\delta k|)$.

In contrast, for the co-propagating crossings in Fig.~\ref{fig:supp_bogoliubov_kinematics}(b), we choose, without loss of generality, the momentum orientation such that both group velocities are positive,
\begin{equation}
E_1(\delta k)\simeq v_1\delta k,
\qquad
E_2(\delta k)\simeq v_2\delta k,
\qquad
v_1,v_2>0.
\label{eq:app_co_propagating_linearization}
\end{equation}
The corresponding factor is
\begin{equation}
\begin{aligned}
\mathcal Q_{\rm co}(\delta k)
&\simeq
\frac{
\Theta(-v_1\delta k)-\Theta(-v_2\delta k)
}{
(v_2-v_1)\delta k
}
\\
&=
0,
\qquad
(\delta k\neq0).
\end{aligned}
\label{eq:app_co_propagating_numerator}
\end{equation}
Thus, for co-propagating states, the two states at a common $\delta k$ lie on the same side of zero energy, as illustrated in Fig.~\ref{fig:supp_bogoliubov_kinematics}(b), and are therefore either both occupied or both empty. Their occupation difference vanishes, so the occupied-to-empty scattering channel responsible for Eq.~\eqref{eq:app_counter_propagating_quotient} is absent. If $v_1=v_2$, both the numerator and denominator vanish within the linearized description; the quotient is then evaluated as the limiting value $-f'(E)$ and remains finite rather than producing a $1/|\delta k|$ Peierls enhancement.

At finite temperature, thermal broadening replaces the zero-temperature $1/|\delta k|$ divergence in Eq.~\eqref{eq:app_counter_propagating_quotient} with a finite response near $\delta k=0$. The underlying kinematic distinction nevertheless remains: opposite group velocities allow the finite-momentum pairing fluctuation to connect occupied and empty low-energy Bogoliubov states, whereas equal-sign velocities do not.

\section{Interaction Dependence of the Superconducting Kohn Anomaly}
\label{supp:interaction_dependence}
\begin{figure}[ht]
    \centering
    \includegraphics[width=0.6\linewidth]{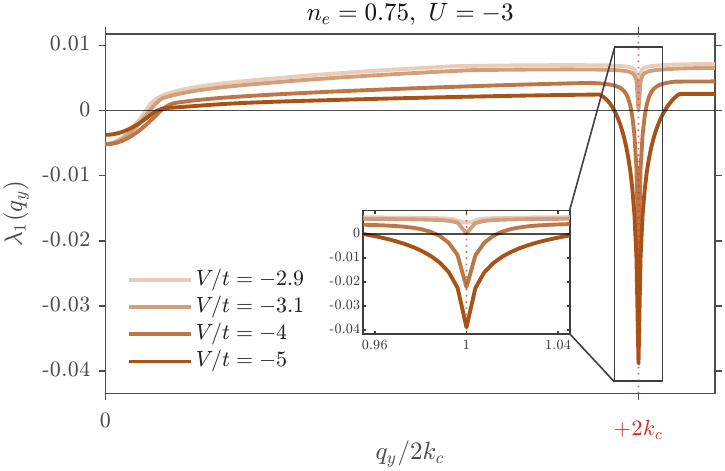}
    \caption{
Nearest-neighbour $V/t$ dependence of the superconducting Kohn anomaly at fixed $n_e=0.75$ and $U/t=-3$. The lowest stiffness eigenvalue $\lambda_1(q_y)$ is shown for $V/t=-2.9,\,-3.1,\,-4$, and $-5$ (increasingly darker colours), for a ribbon with $N_x=20$ and $N_y=N_k=600$. For each value of $V/t$, the momentum axis has been shifted so that the minimum associated with the Kohn anomaly is aligned at $q_y=2k_c$. Inset enlarges the $q_y\simeq2k_c$ region, highlighting the cusp structure of the soft modes.
}
    \label{fig:supp_interaction_kohn_anomaly}
\end{figure}

The main text shows, at the representative point $n_e=0.75$, $U/t=-3$, and $V/t=-4$, that the lowest edge pairing-stiffness eigenvalue develops a negative minimum at the ABS-selected wavevector $q_y=2k_c$, thereby producing the edge PDW. This section establishes how the finite-wavevector Kohn anomaly evolves with interaction strength and distinguishes the appearance of the anomaly from the onset of the edge PDW instability. Figure~\ref{fig:supp_interaction_kohn_anomaly} follows a cut through the $n_e=0.75$ phase diagram at fixed $U/t=-3$ and varying $V/t$. The ABS crossings determine the candidate wavevector $q_y=2k_c$, but their presence alone does not guarantee a superconducting Peierls instability, which occurs only when the quasiparticle response suppresses an edge-localized stiffness eigenvalue enough to make it negative.

The increasingly darker curves in Fig.~\ref{fig:supp_interaction_kohn_anomaly} correspond to stronger nearest-neighbour attraction $|V|/t$ and show the progressive softening of the anomaly. For the weakest attraction shown, $V/t=-2.9$, the stiffness develops a cusp-like minimum at $q_y=2k_c$ but remains positive, indicating a superconducting Kohn anomaly without an ensuing Peierls instability. At stronger attraction, the same finite-wavevector minimum deepens and eventually becomes negative, signalling the onset of the edge PDW instability. Thus, the zero-energy ABS crossings identify the wavevector at which the response is enhanced and the superconducting Kohn anomaly occurs, while the interaction strength determines whether this enhancement is sufficient to generate the edge PDW.

\section{Free-Energy and Self-Consistent BdG Calculations}
\label{supp:free_energy_comparison}

The self-consistent BdG equations, crucial for establishing the edge PDW, admit multiple stationary solutions, so convergence alone does not establish which solution is energetically preferred. We therefore compare converged states at the same electron density using the mean-field Helmholtz free energy. For all results in the main text and in this section, $k_BT/t=10^{-5}$, placing the calculations effectively in the zero-temperature regime on the energy scales relevant here. At fixed particle number, the appropriate thermodynamic potential is the mean-field Helmholtz free energy,
\begin{equation}
F_{\rm MF}=\Omega_{\rm MF}+\mu N_e,
\label{eq:supp_helmholtz_from_grand}
\end{equation}
where $N_e=n_eN_xN_y$. For a converged state, the physical mean-field grand potential, including both redundant spin sectors and the constant mean-field terms, can be written in terms of the positive eigenvalues $E_n$ of the spin-reduced BdG Hamiltonian as~\cite{Hutchinson_2020}
\begin{equation}
\Omega_{\rm MF}
=
\operatorname{Tr}h
-\sum_{E_n>0}E_n
-2k_BT\sum_{E_n>0}\ln\!\left(1+e^{-E_n/(k_BT)}\right)
-\sum_i\frac{|\Delta_0(i)|^2}{U}
-\sum_{i,\delta}\frac{|\Delta_\delta(i)|^2}{V},
\label{eq:supp_mean_field_grand_potential}
\end{equation}
where $h$ is the normal-state particle block of the BdG Hamiltonian and $\delta=\pm\hat{x},\pm\hat{y}$ labels the directed nearest-neighbour pairing fields (see Methods for details). The final two terms are the onsite and bond pairing double-counting corrections due to the mean-field decomposition. For each fixed-density comparison below, Eq.~\eqref{eq:supp_helmholtz_from_grand} is evaluated using the independently determined chemical potential of each self-consistent branch and their common particle number.

\subsection{Fully open-boundary calculations}
\label{supp:fully_open_boundary}

A geometry with open boundary conditions (OBC) in both directions tests whether the edge PDW persists when translation symmetry is not imposed along any boundary and when both edge orientations are present in the same sample. As discussed in Sec.~\ref{app:abs_analytic_estimate}, the edges normal to $x$ are pair-breaking for the $s+d+ip_x$ state, while the edges normal to $y$ are pair-breaking for the $s+d+ip_y$ state.

We use a rectangular $N_x=30$, $N_y=40$ system at $n_e=0.75$ and $U/t=-3$. The unequal edge lengths explicitly distinguish the $s+d+ip_x$ and $s+d+ip_y$ orientations, so their free energies need not be equal. Both orientations can nevertheless be converged as independent self-consistent solutions, where the $s+d+ip_x$ solution develops an edge PDW along the two $x$-normal pair-breaking edges, while the $s+d+ip_y$ solution develops an edge PDW along the two $y$-normal pair-breaking edges.

\begin{figure*}[h!]
\centering
\includegraphics[width=\linewidth]{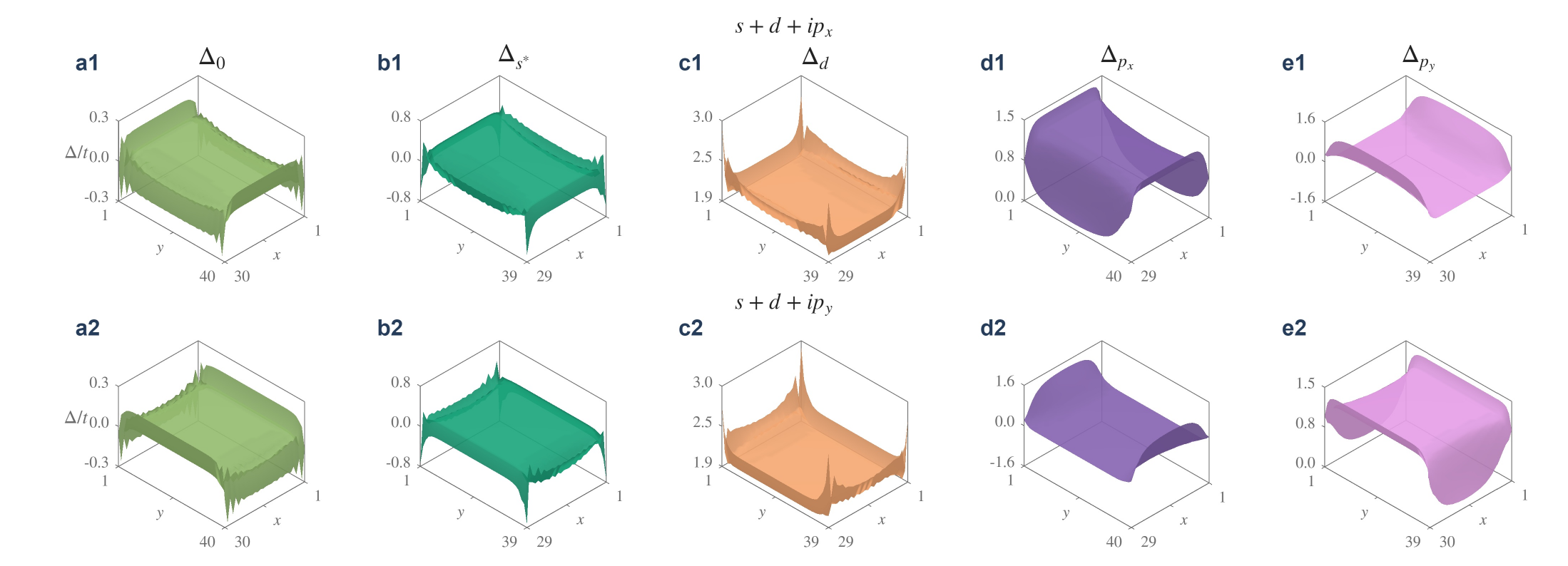}
\caption{Fully open-boundary BdG order parameters for $N_x=30$, $N_y=40$, $n_e=0.75$, $U/t=-3$, and $V/t=-4$. Top row, panels (a1)--(e1), shows the self-consistent $s+d+ip_x$ solution, with the edge PDW along the $x$-normal pair-breaking edges. Bottom row, panels (a2)--(e2), shows the $s+d+ip_y$ solution, with the edge PDW along the $y$-normal pair-breaking edges. Columns display $\Delta_0$, $\Delta_{s^\ast}$, $\Delta_d$, $\Delta_{p_x}$, and $\Delta_{p_y}$ order parameters, respectively.}
\label{fig:supp_obc_pdw_vminus4}
\end{figure*}

We first consider the main-text interaction strength $V/t=-4$. Figure~\ref{fig:supp_obc_pdw_vminus4} shows the finite-$Q_{\rm PDW}$ modulation along the pair-breaking edges of both orientations. The modulation is most clearly resolved in the spin-singlet order parameters, $\Delta_0$, $\Delta_{s^\ast}$, and $\Delta_d$. The $p$-wave components are comparatively smooth and do not exhibit a clearly resolved finite-$Q_{\rm PDW}$ modulation in the real-space surface plots. Fourier analysis finds a weak finite-$Q_{\rm PDW}$ modulation also in these order parameters, but their amplitudes are small compared with the edge modulation of the spin-singlet order parameters. 

\begin{figure*}[h!]
\centering
\includegraphics[width=\linewidth]{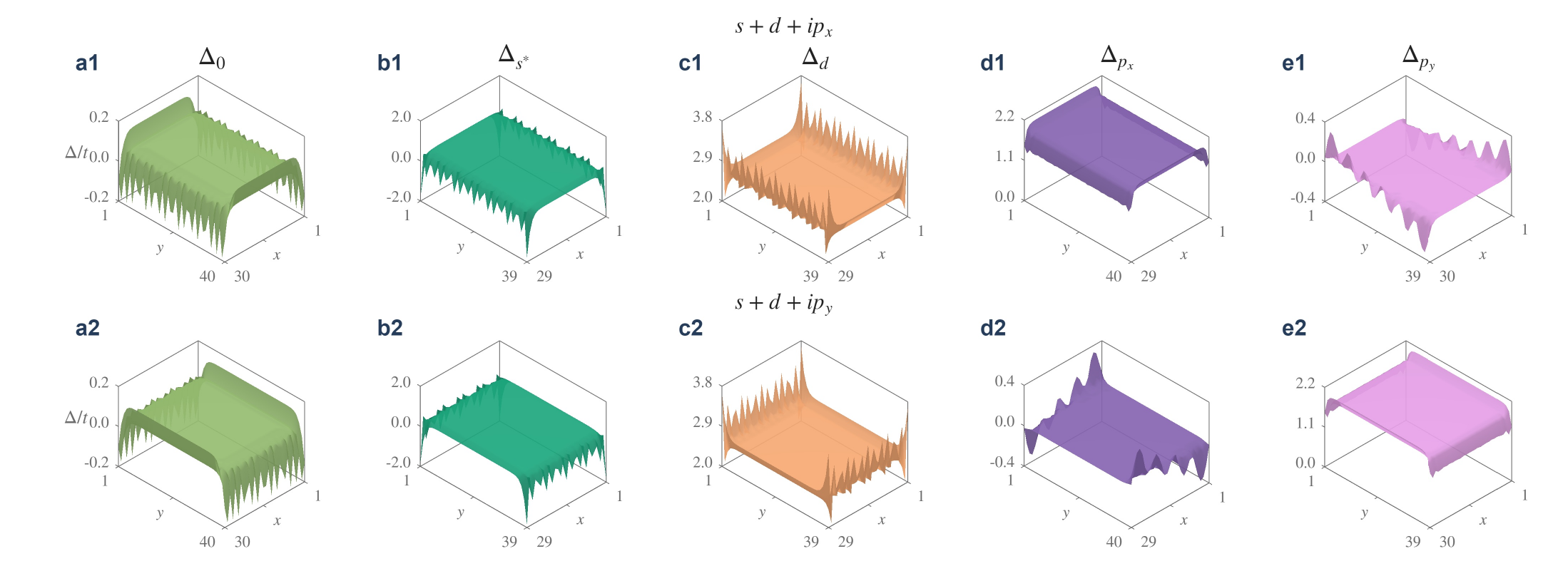}
\caption{Fully open-boundary BdG order parameters for $N_x=30$, $N_y=40$, $n_e=0.75$, $U/t=-3$, and $V/t=-5$. Top row, panels (a1)--(e1), shows the self-consistent $s+d+ip_x$ solution, with the edge PDW along the $x$-normal pair-breaking edges. Bottom row, panels (a2)--(e2), shows the $s+d+ip_y$ solution, with the edge PDW along the $y$-normal pair-breaking edges. The columns display $\Delta_0$, $\Delta_{s^\ast}$, $\Delta_d$, $\Delta_{p_x}$, and $\Delta_{p_y}$ order parameters, respectively.}
\label{fig:supp_obc_pdw_vminus5}
\end{figure*}

To display the edge-PDW modulation more clearly, Fig.~\ref{fig:supp_obc_pdw_vminus5} shows the self-consistent solutions at $V/t=-5$, with all other parameters unchanged. At this stronger nearest-neighbour attraction, the finite-$Q_{\rm PDW}$ modulation is substantially larger, consistent with the deeper negative stiffness minimum in Fig.~\ref{fig:supp_interaction_kohn_anomaly}. Both bulk orientations again develop the edge PDW along their pair-breaking edges.

In addition to the finite-$Q_{\rm PDW}$ edge modulation, panels (e1) and (d2) of Fig.~\ref{fig:supp_obc_pdw_vminus5} show a long-wavelength, sample-scale variation, most visible in $\Delta_{p_y}$ for the $s+d+ip_x$ solution and in $\Delta_{p_x}$ for the $s+d+ip_y$ solution. This component forms a broad, sign-changing background beneath the edge-PDW oscillations and is antisymmetric about both sample midlines, resembling the lowest odd spatial harmonic of the rectangle. We distinguish this secondary modulation from the Peierls-selected edge PDW but do not investigate its microscopic origin further here.

The fixed-density Helmholtz free energies of the four rectangular solutions are listed in Table~\ref{tab:supp_obc_fixed_density_free_energies}. At each interaction strength, the two orientations were converged independently at $n_e=0.75$, allowing their chemical potentials to adjust separately.
\begin{table*}[h!]
\centering
\renewcommand{\arraystretch}{1.15}
\setlength{\tabcolsep}{8pt}
\begin{tabular}{|c|c|c|c|c|}
\hline
$V/t$ & State: $s+d+\Box$ & $\mu/t$ & $n_e$ & $F_{\rm MF}/(N_xN_y t)$ \\
\hline\hline
\multirow{2}{*}{$-4$}
& $ip_x$ & $-0.9047$ & $0.7500$ & $-1.71404$ \\
& $ip_y$ & $-0.9050$ & $0.7500$ & $-1.71532$ \\
\hline
\multirow{2}{*}{$-5$}
& $ip_x$ & $-0.9949$ & $0.7500$ & $-1.83284$ \\
& $ip_y$ & $-0.9967$ & $0.7500$ & $-1.83507$ \\
\hline
\end{tabular}
\caption{Mean-field Helmholtz free energies per lattice site for the fully open-boundary $N_x=30$, $N_y=40$ solutions at $n_e=0.75$, $U/t=-3$, and $k_BT/t=10^{-5}$. In the state column, $\Box=ip_x$ or $ip_y$ specifies the orientation of the mixed-symmetry $s+d+\Box$ solution. Values are rounded to the displayed precision.}
\label{tab:supp_obc_fixed_density_free_energies}
\end{table*}
For the rectangular geometry, the $s+d+ip_y$ solution has the lower free energy at both interaction strengths. Denoting the free energies of the two fully open-boundary orientations by $F_{\text{OBC},\,p_x}$ and $F_{\text{OBC},\,p_y}$, their differences are
\begin{equation}
\frac{F_{\text{OBC},\,p_y}-F_{\text{OBC},\,p_x}}{N_xN_y t}
=
\begin{cases}
-1.28\times10^{-3}, & V/t=-4,\\
-2.23\times10^{-3}, & V/t=-5.
\end{cases}
\label{eq:supp_obc_orientation_free_energy_difference}
\end{equation}
This splitting reflects the explicit breaking of the $90^\circ$ equivalence between the two orientations by the rectangular geometry. Crucially, both self-consistent solutions still develop an edge PDW along their pair-breaking edges, regardless of which orientation has the lower free energy. Therefore, selecting the bulk orientation, as we do in the main text, only changes which pair of edges hosts the PDW, but has no effect on whether the boundary instability occurs. The edge PDW consequently cannot be attributed to the geometry-specific tendency of the ribbon to reorient its bulk order parameter.

\subsection{Ribbon free-energy comparison}

We next consider the ribbon geometry used in the main text, which has open boundaries along $x$ and periodic boundary conditions along $y$. This geometry isolates the pair of edges normal to $x$ while retaining the conserved momentum $k_y$, allowing the ABS zero-energy states and the associated finite-wavevector pairing response to be resolved directly. Because the ribbon contains only this open-edge orientation, it also distinguishes the otherwise symmetry-related $s+d+ip_x$ and $s+d+ip_y$ bulk orientations.

To explicitly establish the energetic stability of the edge PDW solution shown in Fig.~4 of the main text, we compare it directly with the unmodulated ($y$-translation-invariant) $s+d+ip_x$ ribbon solution. For this, we use the same $N_x=20$, $N_y=150$ ribbon at $U/t=-3$ and $V/t=-4$. The main-text edge PDW calculation targeted $n_e=0.75$ and converged to $n_e=0.7478$ within the chosen density tolerance. We take this achieved density as the target for the unmodulated solution and reconverge it with the chemical potential allowed to adjust. Here ``translation invariant'' refers only to the periodic $y$ direction: the order parameters remain free to vary self-consistently across the open $x$ direction. The edge PDW solution is obtained by additionally allowing translation symmetry to break along $y$.

Denoting the Helmholtz free energies of the edge PDW and unmodulated $s+d+ip_x$ solutions by $F_{\rm PDW}$ and $F_{\text{u},\,p_x}$, respectively, we obtain the values listed in Table~\ref{tab:supp_ribbon_pdw_free_energies}.
\begin{table}[ht]
\centering
\renewcommand{\arraystretch}{1.15}
\setlength{\tabcolsep}{8pt}
\begin{tabular}{|l|c|c|c|}
\hline
State & $\mu/t$ & $n_e$ & $F_{\rm MF}/(N_y t)$ \\
\hline\hline
Unmodulated $s+d+ip_x$ & $-0.9119$ & $0.7478$ & $-34.23070$ \\
\hline
Edge-PDW $s+d+ip_x$ & $-0.9120$ & $0.7478$ & $-34.23137$ \\
\hline
\end{tabular}
\caption{Mean-field Helmholtz free energies for the unmodulated and edge-PDW $s+d+ip_x$ solutions on the $N_x=20$, $N_y=150$ ribbon at $U/t=-3$, $V/t=-4$, and $k_BT/t=10^{-5}$. The quantity $F_{\rm MF}/N_y$ is the free energy of the complete ribbon per longitudinal lattice period and includes both open edges. Values are rounded to the displayed precision.}
\label{tab:supp_ribbon_pdw_free_energies}
\end{table}
Allowing translation symmetry to break within the selected $s+d+ip_x$ sector thus lowers the Helmholtz free energy by
\begin{equation}
\frac{F_{\rm PDW}-F_{\text{u},\,p_x}}{N_y t}
=
-6.68\times10^{-4}.
\label{eq:supp_pdw_helmholtz_gain}
\end{equation}
The negative difference confirms that the self-consistent edge PDW is energetically favoured over the unmodulated $s+d+ip_x$ solution.
For completeness, an independently converged $y$-translation-invariant $s+d+ip_y$ ribbon solution has a lower free energy than both these solutions, consistent with the separate $q_y=0$ orientation instability of the selected $s+d+ip_x$ reference state in Fig.~5(a) of the main text. This orientation preference arises because the $x$-normal ribbon contains only one open-edge orientation. The fully open-boundary calculations in Sec.~\ref{supp:fully_open_boundary} instead show that either symmetry-related bulk orientation develops an edge PDW on its pair-breaking edges. The fully open-boundary results therefore establish that the edge PDW persists for either bulk orientation, while the $q_y=0$ ribbon instability is a separate, geometry-specific orientation effect.

\section{Symmetry Structure of the Mixed-Symmetry State and Edge PDW}
\label{app:symmetry_structure}

This section explains the symmetry constraints behind the bulk $s+d+ip$ state and the multicomponent edge PDW found in the ribbon geometry. The first subsection shows how, in the bulk, a pre-existing $d$-wave condensate couples the $p$-wave sector to the two $s$-wave components. The second subsection shows how the same mixed-symmetry structure produces a collective edge fluctuation involving several pairing channels.

\subsection{Symmetry origin of the $d$ to $s+d+ip$ transition}
\label{app:direct_mixed_transition}

Previous studies of the square-lattice extended Hubbard model have found that cooling a $d$-wave state can produce a direct second-order transition
\begin{equation}
  d
  \longrightarrow
  s+d+ip ,
\end{equation}
without an intervening $s+d$ or $d+ip$ phase~\cite{SenarathYapa2025PC_mss,Hutchinson2019}. This simultaneous onset is not obvious from the normal-state symmetry classification, since the two $s$-wave components and the $p$-wave doublet belong to distinct irreducible representations of $C_{4v}$:
\begin{equation}
  \left(\Delta_0,\Delta_{s^\ast}\right)\sim A_1,\qquad
  \Delta_d\sim B_1,\qquad
  \left(\Delta_{p_x},\Delta_{p_y}\right)\sim E .
  \label{eq:app_order_parameter_irreps}
\end{equation}
Here $\Delta_0$ is the onsite $s$-wave order parameter, while $\Delta_{s^\ast}$ is the extended-$s$ component. While they are microscopically distinct, they both transform as the fully symmetric representation $A_1$.

The direct transition becomes natural once the pre-existing $d$-wave condensate is treated as the reference state. The symmetric product of two $p$-wave doublets decomposes as
\begin{equation}
\mathrm{Sym}^2(E)=A_1\oplus B_1\oplus B_2 .
\label{eq:app_sym_square_E}
\end{equation}
Within this decomposition, the quadratic combination
\begin{equation}
Q_{B_1}=\Delta_{p_x}^2-\Delta_{p_y}^2
\label{eq:app_Q_B1}
\end{equation}
transforms as $B_1$. Since $\Delta_d$ also transforms as $B_1$, the product
\begin{equation}
  \Delta_d^\ast Q_{B_1}
  \sim
  B_1\otimes B_1
  =
  A_1
  \label{eq:app_B1_product}
\end{equation}
has the same symmetry as the $s$-wave sector.
Thus, collecting the two $A_1$ components into
\begin{equation}
  \boldsymbol{\Delta}_s
  =
  \begin{pmatrix}
    \Delta_0\\
    \Delta_{s^\ast}
  \end{pmatrix},
\end{equation}
the Landau free energy admits the gauge-invariant coupling
\begin{equation}
  f_{\rm mix}
  =
  \boldsymbol{\Delta}_s^\dagger
  \boldsymbol{\lambda}
  \Delta_d^\ast (\Delta_{p_x}^2-\Delta_{p_y}^2)
  +
  \mathrm{c.c.},
  \label{eq:app_two_s_mixing}
\end{equation}
where
\begin{equation}
  \boldsymbol{\lambda}
  =
  \begin{pmatrix}
    \lambda_0\\
    \lambda_{s^\ast}
  \end{pmatrix}.
\end{equation}
Gauge invariance follows because every pairing field transforms as $\Delta_\Box\rightarrow e^{i\varphi}\Delta_\Box$, leaving $\boldsymbol{\Delta}_s^\dagger\Delta_d^\ast Q_{B_1}$ invariant.

Now consider a reference $d$-wave state with $\Delta_d=\bar{\Delta}_d\neq0$. The part of the free energy involving the two $A_1$ components is
\begin{equation}
\begin{aligned}
f_{A_1}
&=
\boldsymbol{\Delta}_s^\dagger\mathcal A_s\boldsymbol{\Delta}_s
+
\left[
\boldsymbol{\Delta}_s^\dagger\boldsymbol{\lambda}\bar{\Delta}_d^\ast
\left(
\Delta_{p_x}^2-\Delta_{p_y}^2
\right)
+
\mathrm{c.c.}
\right],
\end{aligned}
\label{eq:app_A1_free_energy}
\end{equation}
where $\mathcal A_s$ is the Hermitian quadratic kernel of the two $s$-wave components. We first assume that the $A_1$ sector remains noncritical at the $d\rightarrow s+d+ip$ transition, so that this transition is driven by the $p$-wave sector. The matrix $\mathcal A_s$ is then positive definite at the transition, and the $s$-wave components can be integrated out. Minimizing over $\boldsymbol{\Delta}_s$ gives
\begin{equation}
\boldsymbol{\Delta}_s
=
-
\mathcal A_s^{-1}
\boldsymbol{\lambda}
\bar{\Delta}_d^\ast
\left(
\Delta_{p_x}^2-\Delta_{p_y}^2
\right)
+
O(|\Delta_p|^4).
\label{eq:app_induced_s_vector}
\end{equation}
The onset of $p$-wave order therefore acts through the pre-existing $d$-wave condensate as a source for both $s$-wave components. Thus, away from special points in parameter space, both $\Delta_0$ and $\Delta_{s^\ast}$ are induced as soon as a $p$-wave component becomes nonzero. In this description, the $p$-wave order is the primary critical field driving the $d\rightarrow s+d+ip$ transition, while $\Delta_0$ and $\Delta_{s^\ast}$ are secondary order parameters.

For a continuous $d\rightarrow s+d+ip$ transition at $T_{c2}$, the primary and secondary components scale as
\begin{align}
|\Delta_p|
&\propto
(T_{c2}-T)^{1/2},
\\
|\Delta_0|,\,
|\Delta_{s^\ast}|
&\propto
|\Delta_p|^2
\propto
T_{c2}-T .
\label{eq:app_critical_scaling}
\end{align}
The $s$- and $p$-wave components can therefore appear at the same temperature $T_{c2}$ without requiring an accidental degeneracy between the $A_1$ and $E$ channels, although they exhibit different critical scaling. The relative $\pm i$ phase between the singlet and triplet sectors is selected by phase-sensitive quartic terms together with the unitarity condition; it is not fixed by the irreducible-representation labels alone.

If the $A_1$ sector instead becomes critical while the $p$-wave sector remains noncritical, the coefficients realized in the microscopic model favour a $d\rightarrow s+id$ transition. The relative $\pm\pi/2$ phase between the $s$- and $d$-wave components is selected by their phase-sensitive quartic coupling; it is not fixed by symmetry alone. Because the coupling in Eq.~\eqref{eq:app_two_s_mixing} is quadratic in the $p$-wave fields, the newly developed $s$-wave order does not induce $\Delta_{p_x}$ or $\Delta_{p_y}$ linearly at the $d\rightarrow s+id$ transition. In the resulting $s+id$ state, however, the same coupling modifies the quadratic form of the $p$-wave sector and may drive a separate $p$-wave instability at a lower temperature. The symmetry and relative phases of the resulting coexistence state depend on the remaining terms in the Landau free energy and cannot be determined from Eq.~\eqref{eq:app_two_s_mixing} alone. In particular, reaching the $s+d+ip$ state from the $s+id$ state would require a corresponding rearrangement of the relative $s$--$d$ phase. A simultaneous continuous onset of the $s$- and $p$-wave components when the $A_1$ sector is itself critical would require coincident $A_1$ and $E$ instabilities; alternatively, a direct first-order $d\rightarrow s+d+ip$ transition could produce their simultaneous onset. The direct continuous $d\rightarrow s+d+ip$ transition found in Ref.~\cite{SenarathYapa2025PC_mss} corresponds to the case analyzed above, in which the $p$-wave sector is primary and the two $s$-wave components are induced as secondary order parameters.

\subsection{Multicomponent structure of the edge pair-density wave}
\label{app:edge_pdw_multicomponent}

This subsection explains why the softened edge pairing mode in Fig.~5(b) of the main text involves several pairing channels and why the self-consistent edge PDW in Fig.~4 additionally contains a spatially shifted $\Delta_{p_y}$ modulation. Near its onset, the edge PDW can be analyzed as a finite-momentum fluctuation about the $y$-translation-invariant $s+d+ip_x$ ribbon state. For a ribbon open along $x$ and periodic along $y$, we write
\begin{equation}
  \Delta_\Box(x,y)=\bar{\Delta}_\Box(x)+\delta\Delta_\Box(x,y),
  \label{eq:app_edge_fluctuation_expansion}
\end{equation}
where $\Box\in\{0,s^\ast,d,p_x,p_y\}$ labels the five pairing channels. The translation-invariant state satisfies
\begin{equation}
  \bar{\Delta}_0,\,\bar{\Delta}_{s^\ast},\,\bar{\Delta}_d,\,\bar{\Delta}_{p_x}\neq0,\qquad \bar{\Delta}_{p_y}=0 .
  \label{eq:app_uniform_parent}
\end{equation}
Although this state is translation invariant along $y$, the profiles $\bar{\Delta}_\Box(x)$ vary self-consistently across the ribbon and include the static reconstruction near the open boundary. Because $\bar{\Delta}_{p_x}\neq0$, the invariant in Eq.~\eqref{eq:app_two_s_mixing} generates bilinear mixing between fluctuations. Expanding to quadratic order gives representative terms
\begin{equation}
  \delta f_{\rm mix}^{(2)}
  \supset{}
  2\lambda_0\bar{\Delta}_{p_x}
  \left(
  \bar{\Delta}_d^\ast
  \delta\Delta_0^\ast
  \delta\Delta_{p_x}
  +
  \bar{\Delta}_0^\ast
  \delta\Delta_d^\ast
  \delta\Delta_{p_x}
  \right)
  +
  2\lambda_{s^\ast}\bar{\Delta}_{p_x}
  \left(
  \bar{\Delta}_d^\ast
  \delta\Delta_{s^\ast}^\ast
  \delta\Delta_{p_x}
  +
  \bar{\Delta}_{s^\ast}^\ast
  \delta\Delta_d^\ast
  \delta\Delta_{p_x}
  \right)
  +
  \mathrm{c.c.}
  \label{eq:app_pdw_channel_mixing}
\end{equation}
Thus the fluctuations $\delta\Delta_0$, $\delta\Delta_{s^\ast}$, and $\delta\Delta_d$ are bilinearly coupled to $\delta\Delta_{p_x}$ in the mixed-symmetry state. These four channels therefore form a collective fluctuation sector rather than independent single-channel instabilities, consistent with the multicomponent softened eigenvector in Fig.~5(b) of the main text.

The fifth fluctuation component, $\delta\Delta_{p_y}$, enters the edge PDW through a different route. Although it is included in the full quadratic fluctuation basis, its weight in the softened eigenvector in Fig.~5(b) of the main text is negligible, showing that it is not a primary component of the instability. Nevertheless, a secondary $\delta\Delta_{p_y}$ modulation is induced in the finite-amplitude self-consistent PDW shown in Fig.~4 of the main text. The local invariant in Eq.~\eqref{eq:app_two_s_mixing} can renormalize the quadratic stiffness of $\delta\Delta_{p_y}$, but it does not generate bilinear mixing between $\delta\Delta_{p_y}$ and the other fluctuation channels because $\bar{\Delta}_{p_y}=0$. We now show that this secondary modulation is instead induced by a symmetry-allowed edge-gradient coupling. At an edge normal to $x$, the remaining reflection symmetry is
\begin{equation}
  M_y:\quad y\rightarrow -y ,
\end{equation}
and components $\Delta_0$, $\Delta_{s^\ast}$, $\Delta_d$, and $\Delta_{p_x}$ are even under $M_y$, while both $\Delta_{p_y}$ and $\partial_y$ are odd. Hence $\partial_y\Delta_{p_y}=\partial_y\delta\Delta_{p_y}$ is even under $M_y$, and the edge permits the quadratic Lifshitz-type mixing
\begin{equation}
\delta f_{\rm L}^{(2)}
=
\sum_{\zeta}
\left[
\kappa_\zeta(x)\delta\Delta_\zeta^\ast\partial_y\delta\Delta_{p_y}
+
\mathrm{c.c.}
\right].
\label{eq:app_edge_lifshitz}
\end{equation}
Here $\zeta\in\{0,s^\ast,d,p_x\}$ labels the fluctuation components that are even under $M_y$. The functions $\kappa_\zeta(x)$ are edge-localized coupling coefficients, distinct from the transverse fluctuation profiles $w_\ell(x)$ used in the stiffness calculation. Fourier transforming Eq.~\eqref{eq:app_edge_lifshitz} along the edge replaces $\partial_y$ by $iq_y$, giving contributions proportional to
\begin{equation}
\sum_\zeta\left[
iq_y\kappa_\zeta(x)\delta\Delta_{\zeta,q_y}^\ast(x)\delta\Delta_{p_y,q_y}(x)
+
\mathrm{c.c.}
\right].
\end{equation}
The Lifshitz coupling therefore vanishes at $q_y=0$ and mixes $\delta\Delta_{p_y,q_y}$ with the even fluctuation components only at finite momentum. A real standing-wave PDW is formed by combining the symmetry-related components at $\pm Q_{\rm PDW}$. Choosing the spatial origin so that the $M_y$-even components are cosine-like, reflection symmetry requires the $M_y$-odd $\Delta_{p_y}$ component to be sine-like:
\begin{equation}
\begin{aligned}
\delta\Delta_\zeta(x,y)&=A_\zeta(x)\cos(Q_{\rm PDW}y),\\
\delta\Delta_{p_y}(x,y)&=A_{p_y}(x)\sin(Q_{\rm PDW}y).
\end{aligned}
\label{eq:app_even_odd_pdw_components}
\end{equation}
Substituting Eq.~\eqref{eq:app_even_odd_pdw_components} into Eq.~\eqref{eq:app_edge_lifshitz} shows why this cosine--sine structure is energetically coupled. The derivative of the $p_y$ component is
\begin{equation*}
\partial_y\delta\Delta_{p_y}(x,y)=Q_{\rm PDW}A_{p_y}(x)\cos(Q_{\rm PDW}y),
\end{equation*}
which has the same spatial dependence as the even components. Their product therefore has a nonzero average along the edge and contributes to the Lifshitz energy. By contrast, if $\delta\Delta_{p_y}$ were cosine-like and spatially in phase with the even components, its derivative would be sine-like and the product would average to zero over a modulation period. The Lifshitz coupling therefore favours a $\pi/2$ spatial shift between $\Delta_{p_y}$ and the even components, with the sign of the shift determined by the coupling coefficients $\kappa_\zeta(x)$. This is a phase shift of the modulation along the edge, distinct from the internal complex phase differences among the superconducting order parameters.

This structure agrees with the self-consistent BdG solution in Fig.~4(a2)--(e2) of the main text: the dominant even components oscillate approximately in phase, while the induced $\Delta_{p_y}$ component is shifted by $\pi/2$. Thus, the edge-allowed Lifshitz coupling selects the cosine--sine structure of the edge PDW, in which the induced $\Delta_{p_y}$ modulation is sine-like relative to the cosine-like dominant components.

\section{Projection of the Finite-Momentum Pair Field onto the ABS Subspace}
\label{app:abs_projection_pair_field}

This section shows explicitly how the gap-opening quasiparticle expectation value highlighted in Eq.~(9) of the main text becomes a finite-momentum Cooper-pair amplitude in the microscopic electron basis. We use the momentum-space spin-reduced Nambu spinor $\Psi_{k_y}$ defined in Eq.~(D.2) of Methods. The translation-invariant ribbon Hamiltonian, given in Eq.~(D.3) of Methods, is repeated here for convenience:
\begin{equation}
\hham_0=\sum_{k_y}\Psi^\dagger_{k_y}\mathcal H_{\rm BdG}(k_y)\Psi_{k_y}+E_c.
\label{eq:supp_uniform_bdg_hamiltonian}
\end{equation}
The Bogoliubov mode operators are defined from the corresponding eigenstates by
\begin{equation}
\mathcal H_{\rm BdG}(k_y)|n,k_y\rangle=E_n(k_y)|n,k_y\rangle,\qquad \hat\gamma_{n,k_y}=\langle n,k_y|\Psi_{k_y}.
\label{eq:app_gamma_definition}
\end{equation}

At the mean-field level, a finite-momentum pairing deformation enters the BdG Hamiltonian through terms that create or annihilate an electron pair carrying momentum $q_y$, and is therefore bilinear in the electron operators. Because each electron operator is a superposition of Bogoliubov creation and annihilation operators, the same deformation generally produces terms of the forms $\hat\gamma^\dagger\hat\gamma$, $\hat\gamma^\dagger\hat\gamma^\dagger$, and $\hat\gamma\hat\gamma$ in the Bogoliubov basis. The $\hat\gamma^\dagger\hat\gamma$ terms describe scattering between Bogoliubov states, while the other two terms create or annihilate pairs of Bogoliubov quasiparticles. We are interested in the finite-momentum scattering between the two zero-energy states $|{\rm ABS},\pm k_c\rangle$, because it mixes these states and opens the ABS gap. We therefore retain only these two low-energy states; this restriction is what we mean by projection onto the ABS subspace. We suppress the ABS band label below and write $\hat\gamma_k\equiv\hat\gamma_{{\rm ABS},k}$.

Denoting the amplitude of this gap-opening scattering process by $\Delta_{\rm PDW}$, the term introduced in Eq.~(8) of the main text is
\begin{equation}
\hham_{\rm PDW}^{\rm MF}\supset\Delta_{\rm PDW}\hat\gamma^\dagger_{k_c}\hat\gamma_{-k_c}+\Delta_{\rm PDW}^{\ast}\hat\gamma^\dagger_{-k_c}\hat\gamma_{k_c}.
\label{eq:app_abs_gap_hamiltonian}
\end{equation}
The gap-opening mean-field channel may be written schematically as
\begin{equation}
\Delta_{\rm PDW}\propto\left\langle\hat\gamma^\dagger_{-k_c}\hat\gamma_{k_c}\right\rangle.
\label{eq:app_abs_gap_mean_field}
\end{equation}

To see how this expectation value contains a finite-momentum pair amplitude in the microscopic electron basis, we write the ABS operator as
\begin{equation}
\hat\gamma_k=u_k^\ast\hat c_k+v_k^\ast\hat c^\dagger_{-k},
\label{eq:app_gamma_schematic}
\end{equation}
where $u_k$ and $v_k$ denote the electron and hole components of the ABS eigenstate $|{\rm ABS},k\rangle$, with the transverse and spin indices suppressed. The ABS expectation value then decomposes as
\begin{equation}
\left\langle\hat\gamma^\dagger_{-k_c}\hat\gamma_{k_c}\right\rangle
={}
u_{-k_c}u_{k_c}^{\ast}
\left\langle\hat c^\dagger_{-k_c}\hat c_{k_c}\right\rangle
+
u_{-k_c}v_{k_c}^{\ast}
\left\langle\hat c^\dagger_{-k_c}\hat c^\dagger_{-k_c}\right\rangle
+
v_{-k_c}u_{k_c}^{\ast}
\left\langle\hat c_{k_c}\hat c_{k_c}\right\rangle
+
v_{-k_c}v_{k_c}^{\ast}
\left\langle\hat c_{k_c}\hat c^\dagger_{-k_c}\right\rangle .
\label{eq:app_abs_coherence_full_decomposition}
\end{equation}
The second and third terms are anomalous in the microscopic electron basis and correspond to Cooper-pair amplitudes with center-of-mass momenta $-2k_c$ and $+2k_c$, respectively.

The electron operators in Eq.~\eqref{eq:app_abs_coherence_full_decomposition} are written schematically. Restoring the transverse coordinate and the $S_z=0$ spin structure, the anomalous contributions take the form
\begin{equation}
\sum_{xx'}\mathcal W^{(-)}_{xx'}\left\langle\hat c^\dagger_{x,-k_c,\uparrow}\hat c^\dagger_{x',-k_c,\downarrow}\right\rangle
+
\sum_{xx'}\mathcal W^{(+)}_{xx'}\left\langle\hat c_{x,k_c,\downarrow}\hat c_{x',k_c,\uparrow}\right\rangle,
\label{eq:app_abs_pair_terms_indexed}
\end{equation}
where the weights $\mathcal W^{(\pm)}_{xx'}$ are fixed by the electron and hole components of the corresponding ABS wave functions. Fermionic antisymmetry fixes the exchanged spin components; for example,
\begin{equation}
\left\langle\hat c^\dagger_{x,-k_c,\uparrow}\hat c^\dagger_{x',-k_c,\downarrow}\right\rangle
=
-
\left\langle\hat c^\dagger_{x',-k_c,\downarrow}\hat c^\dagger_{x,-k_c,\uparrow}\right\rangle.
\end{equation}
The two terms in Eq.~\eqref{eq:app_abs_pair_terms_indexed} carry center-of-mass momenta $-2k_c$ and $+2k_c$, respectively. Thus the same gap-opening ABS expectation value in Eq.~\eqref{eq:app_abs_gap_mean_field} contains finite-momentum Cooper-pair amplitudes in the electron basis, which are precisely the microscopic components of the edge PDW described in the main text.

\end{document}